\RequirePackage{fix-cm}
\documentclass[twocolumn]{svjour3}          % twocolumn
\usepackage[linesnumbered, ruled, vlined]{algorithm2e}
\usepackage{xcolor}
\usepackage{pgfplots}
\usepackage{caption}
\usepackage{subcaption}
\usepackage{float}
\usepackage{stfloats}
\usepackage[normalem]{ulem}
\usepackage{multirow}
\usetikzlibrary{pgfplots.groupplots}
\usepackage{tabularx}
\usepackage{adjustbox}
\usepackage[noadjust]{cite}
\usepackage{graphicx}
\usepackage{graphics}
\usepackage{url}

\SetKwRepeat{Do}{do}{while}%

\definecolor{RYB1}{RGB}{141, 211, 199}
\definecolor{RYB2}{RGB}{244, 222, 33}
\definecolor{RYB3}{RGB}{190, 186, 218}
\definecolor{RYB4}{RGB}{251, 128, 114}
\definecolor{RYB5}{RGB}{128, 177, 211}
\definecolor{RYB6}{RGB}{253, 180, 98}
\definecolor{RYB7}{RGB}{179, 222, 105}
\definecolor{RYB8}{RGB}{199, 21, 133}

\pgfplotscreateplotcyclelist{colorbrewer-RYB}{
{RYB1!50!black,fill=RYB1},
{RYB2!50!black,fill=RYB2},
{RYB3!50!black,fill=RYB3},
{RYB4!50!black,fill=RYB4},
{RYB5!50!black,fill=RYB5},
{RYB6!50!black,fill=RYB6},
{RYB7!50!black,fill=RYB7},
{RYB8!50!black,fill=RYB8},
}

\pgfplotsset{
    select row/.style={
        x filter/.code={\ifnum\coordindex=#1\else\fi}
    }
}

\smartqed  % flush right qed marks, e.g. at end of proof
\usepackage{graphicx}
\begin{document}

\title{
% The 
% Generic Framework 
% of Update Memo technique for 
% various 
% LSM secondary indexes

% 1. Supporting Update-intensive LSM secondary indexes
% workloads for LSM secondary indexes
% Update-intensive LSM Secondary Indexes
An Update-intensive LSM-based R-tree  %RUM-tree
Index
%WGA: I changed the title a bit. Let me know if you disagree with it.
}

% \subtitle{Do you have a subtitle?\\ If so, write it here}


\author{Jaewoo Shin         \and
        Jianguo Wang        \and
        Walid G. Aref
}


\institute{J. Shin \at
              Purdue University, West Lafayette, IN \\
              \email{shin152@purdue.edu}           %  \\
           \and
           J. Wang \at
              Purdue University, West Lafayette, IN \\
              \email{csjgwang@purdue.edu}           %  \\
           \and
           W. G. Aref  \at
              Purdue University, West Lafayette, IN \\
              \email{aref@purdue.edu}           %  \\
}

\date{Received: date / Accepted: date}

\maketitle
% page numbers
\thispagestyle{plain}
\pagestyle{plain}

\begin{abstract}
\begin{sloppypar}
Many applications require update-intensive workloads on spatial objects, e.g., social-network services and shared-riding services that track moving objects. %(devices).
By buffering insert and delete operations in memory, the Log Structured Merge Tree (LSM) has been used widely in various systems because of its ability to handle 
%update-intensive
write-heavy
workloads. 
While the focus on LSM has been on key-value stores and their optimizations, 
there is a need to study how to efficiently support LSM-based {\em secondary} indexes (e.g., location-based indexes) as modern, heterogeneous data necessitates the use of secondary indexes.
%\walid{can you state why this problem is difficult or nontrivial before you suggest the augmentation of a memo (which is the solution to some hard challenge that you need to highlight here)?}\jaewoo{Edited}
In this paper, we investigate the augmentation of a main-memory-based memo structure into an LSM secondary index structure to handle update-intensive workloads efficiently. 
We conduct this study in the context of an R-tree-based secondary index. In particular,
we introduce the LSM RUM-tree that demonstrates the use of an Update Memo in an LSM-based R-tree to enhance the  performance of the R-tree's insert, delete, update, and search operations.
The LSM RUM-tree introduces 
%novel 
new
strategies to control the size of the Update Memo to make sure it always fits in memory for high performance. 
The Update Memo is a light-weight in-memory structure that is suitable for handling update-intensive workloads without introducing significant overhead. 
Experimental results using real spatial data demonstrate that the LSM RUM-tree achieves up to 9.6x speedup on update operations and up to 2400x speedup on query processing over existing LSM R-tree implementations.
\end{sloppypar}

%WGA: Check the above key words if they are OK or not.
\end{abstract}

\section{Introduction}

%WGA: MOVE the sentence below out of the abstract  into the introduction
%JS: Done, here.
% This paper is an extension over our previous work on the LSM RUM-tree,
% %\cite{Add citation}
% where we deliver in-depth analysis of the improvement in performance. %improvement. 
% In addition, we %endeavor to include 
% introduce 
% concurrency support %of 
% to
% the Update Memo, where the experiments show 4.5x speedup on concurrent update operations over the existing and baseline implementations. 
% \jianguo{Is this paragraph supposed to be here? We can delete it because we have mentioned it at the end of introduction.}\jaewoo{modified/moved to the end}

\begin{sloppypar}
In recent years, massive amounts of location data have been generated continuously from mobile devices, social media, and shared-riding services. As devices or objects move in space, they update their locations and expect to have responsive services (e.g., getting weather emergencies and location-targeted advertisements). From a system's perspective, it is challenging to efficiently handle update-intensive location workloads, and answer queries 
with low latency.
\end{sloppypar}

A widely-used approach for write-intensive workloads is the Log-Structured Merge tree (or LSM, for short)~\cite{o1996log}. 
The main idea of LSM is to buffer data ingestion in memory, and then periodically flush the buffers into disk. This can convert random I/Os to sequential I/Os for heavy write workloads. The LSM R-tree has been proposed to handle write-intensive spatial workloads~\cite{alsubaiee2014storage}. Locations or coordinates are not a primary key, but rather a secondary key. 
The reason is that  location data cannot uniquely identify an object in the same way an object identifier does. Moreover, a certain location can be shared by multiple objects at different times. 

Determining the most recent state of an object (in this case, the object's current location)  in  write-intensive (update-intensive) workloads without much overhead is challenging. For example, consider 
a moving-objects spatial database application that maintains the current locations of objects moving in space.
This application is update-intensive due to the continuous movements of the objects. An LSM-like structure for storing location data would be a perfect match. 
The LSM R-tree~\cite{alsubaiee2014storage, luo2019efficient} has been introduced as a way to ingest object movements in space while at the same time being able to answer location-oriented  queries, e.g., range queries. In the LSM R-tree,  an object moving in space continuously updates the object's location into an in-memory R-tree. There is only one ``current" location for an object, and all the previous locations for the object become outdated. When the allocated memory for the in-memory R-tree runs out, it is dumped to disk, and a new in-memory R-tree is started.
When querying the R-tree, e.g., a range search query, all the in-memory and disk-based R-trees are probed, and hence the results may contain outdated locations of the object. Unless handled properly, this could lead to incorrect answers to the range query. Thus, additional measures are taken in the LSM R-tree to ensure correct results~\cite{alsubaiee2014storage, luo2019efficient}.

Alsubaiee et al.~\cite{alsubaiee2014storage} address this issue 
by using an {\em eager} strategy, where an additional data structure, namely, a deleted-key B$^+$-tree, is associated with an in-memory R-tree to store the deleted objects' keys. This  indicates that an old version of the object is deleted, and a new one is inserted. The deleted-key B$^+$-tree is also flushed to disk along with the corresponding R-tree. This scheme induces extra overhead due to the need to maintain and access the deleted-key B$^+$-tree, and thus affects 
%negatively 
both the update and query performances.

Another solution to address the issue is to use a {\em validation} strategy~\cite{luo2019efficient}, where the deleted-key B$^+$-tree that is coupled with each R-tree is removed. Instead, a primary key index (i.e., a B$^+$-tree holding the primary keys) with timestamps is used to lazily clean the obsolete objects during the merge operation of the LSM-based secondary index.
The decoupled primary key index with timestamps helps avoid the extra maintenance cost, and shows improvement in update performance. However, this approach induces extra overhead in the search operation to validate the most recent state of an object.
While enhancing  update performance, this approach 
%worsens the 
penalizes
search in contrast to the eager strategy. 

%On the other side, 
The 
% RUM-tree~\cite{xiong2006r} 
RUM-tree~\cite{silva2009rum,xiong2006r} 
%WGA: Add the VLDB Journal reference here
%JS: Done
has been introduced where a disk-based R-tree handles frequently-updated location data 
using
an in-memory structure, termed the Update Memo (UM), by maintaining the timestamp of the most-recent object-update in the UM. 
The UM has been utilized in the context of the R-tree~\cite{xiong2006r,silva2009rum,Zhu2013WAIM}, the Grid File~\cite{xiong2006lugrid}, in indexing limited trajectories~\cite{AhmedTSAS2018}, and in indexing current and near-future locations of moving objects~\cite{chen2008benchmark}.
Because the UM keeps every 
%updates 
update
in memory, its size 
%is matter. 
matters.
In the 
% RUM-tree~\cite{xiong2006r},
RUM-tree~\cite{xiong2006r,silva2009rum} 
%WGA: Add the VLDB Journal reference here
%JS: Done.
the UM size is 
controlled (upper-bounded)
%shrunk 
by probing the R-tree nodes in disk 
to lazily perform garbage cleaning. However, garbage cleaning strategies in the original RUM-tree and its variants are not directly applicable to an LSM-based R-tree index due to the fact that in LSM, multiple R-trees 
%end up being dumped to the 
exist as 
disk components while having stored in them some of the obsolete (out-dated) locations of the moved objects.
In this case,  garbage cleaning would become expensive if applied without adaptation to the LSM environment.
It is 
non-trivial to get rid of these out-dated objects in the R-tree, and thus it is more challenging to
enforce an upper-bound on
the size of the UM when  deployed in an LSM-based index. Moreover, the original Update Memo is sensitive to the insert commands, i.e., it gets updated when an insert operation is performed on the accompanying index. This is not practical especially for insert-intensive workloads. 
Even if we maintain multiple deleted-key B$^+$-trees~\cite{alsubaiee2014storage} or a primary key index~\cite{luo2019efficient} in the memory layer, 
that would not be a sufficient solution for update-intensive workloads. The reason is that having multiple deleted-key B$^+$-trees
would require
%lead to 
multi-path searches in each deleted-key B$^+$-tree,  and the maintenance cost for the orchestration 
with the LSM R-tree 
% the corresponding
%WGA: is adding the word "the corresponding" make sense?
%JS: deleting the word looks better.
%will 
would
still exist. Similarly, having a  primary key index in memory
  does not scale. The reason is that  as it may exceed the memory budget because the size of the primary key index is proportional to the total number of primary keys, which is not guaranteed to fit in memory.
  
  We propose a new scheme that improves 
  %the  
  both the update and search performances. We adapt and investigate the use of the Update Memo in the context of an LSM secondary index, mainly the R-tree. We refer to the new UM-enabled index by the LSM RUM-tree.  

%\vspace{10pt}
 The main contributions of this paper are as follows:
\begin{itemize}
    \item We introduce the LSM RUM-Tree, an LSM-based R-tree that utilizes an in-memory Update Memo (UM) to support update-intensive spatial workloads  without sacrificing 
    search performance. The LSM RUM-Tree maintains multiple R-trees on disk but its UM component that resides only in memory is bounded in size (not proportional to the number of objects in the system). The UM is guaranteed not to exceed its allocated memory. 
    \item The LSM RUM-Tree is presented in the context of geo-location data. However, the  LSM-based Update Memo (UM) can be applied to other update-intensive secondary indexes, e.g., a B$^+$-tree or an inverted index 
    (Section~\ref{sec:lsmrum}).
    \item We introduce four cleaning strategies to bound the UM size in memory and the R-trees in disk. This does not only reduce the consumed memory space, but also improves the overall performance (Section~\ref{sec:cleaning}). 
    \item We conduct extensive experiments using real datasets to replicate update-intensive workloads. The experimental results demonstrate that the LSM RUM-tree  enhances the performance of the LSM-based R-tree on the delete,  update, and search operations.
    The LSM RUM-tree achieves up to 9.6x speedup on update operations and up to 2400x speedup in search performance over state-of-the-art LSM R-tree implementations.
    % The LSM RUM-tree is implemented inside  AsterixDB~\cite{asterixdb}. The LSM RUM-tree is open-sourced at \textcolor{blue}{https://github.com/nujwoo/lsmrum}.  ~(Section~\ref{sec:performance}).
\end{itemize}

% \jianguo{Add section numbers for each of the following item. Cite correctly for the LSM RUM-tree.} \jaewoo{added}
This paper is an extended version of a short paper that introduces the LSM RUM-tree~\cite{shin2021lsm}. This paper extends over the short paper~\cite{shin2021lsm} in the following ways.
%, where it focuses mainly on the Update Memo technique for LSM R-tree (LSM RUM-tree) in a single threaded environment. In this paper, however, we provide a generic framework of the Update Memo technique and discuss how it works in multi-thread environment. The additional contribution of this paper are as follows:

\begin{itemize}
    % \item We extend previous work on Update Memo technique. We provide a generic framework of the Update Memo to be orchestrated with any other LSM secondary indexes, such as LSM B-tree and/or LSM inverted index. We present how the Update Memo could be used with other indexes and discuss their performance improvements. 
    %WGA: These contributions are not well phrased. Please make sure to write them better.
    \item 
We extend the analysis of LSM-aware UM cleaning strategies to show how well each of the UM cleaning strategies shrinks the UM size, and achieves performance improvements (Section~\ref{sec:cleaning}).
    \item We 
    introduce  concurrency control support for the Update Memo. 
    This is essential to support concurrent activities over the LSM RUM-Tree, e.g., in a multi-threaded/multi-core setup. 
    We introduce the atomic operation {\em Compare and If Less then Swap} (CILS) to facilitate concurrent activities on the UM in a multi-threaded environment (Section~\ref{sec:concurrency}). The experiments show 4.5x speedup on concurrent update operations over the existing and baseline implementations.
    % \item We provide in-depth analysis of the LSM RUM-tree to show where the performance improvement is achieved over the existing implementations. 
    \item Our comprehensive analysis of the LSM RUM-tree aims to demonstrate where performance improvements have been made over existing implementations.
%\walid{I have a problem with your saying: to show "where" What do you mean by "where"? Try to rephrase.} \jaewoo{how about now?}
    We include new additional LSM RUM-tree experiments over the ones in the original short paper, and discuss the experimental results (Section~\ref{sec:performance}). 
%    \walid{Can you say how many new experiments have you added, e.g., 10, 15, etc.?} \jaewoo{We've added 2 more detailed results (tables) and 4 more performance experiments per dataset}
    % \item The LSM RUM-tree is implemented inside AsterixDB~\cite{asterixdb} with concurrency support, and we experimentally evaluate the performance improvements, and discuss where the improvements are achieved in detail.
\end{itemize}

The rest of this paper proceeds as follows. Section~\ref{sec:bg} presents  background material and discusses the related work.  Section~\ref{sec:lsmrum} introduces the LSM RUM-tree, all its supporting index operations, and shows how the Update Memo (UM) is adapted  for use inside the LSM RUM-tree. Section~\ref{sec:cleaning} presents how the cleaning strategies bound the size of the UM, and improve the performance of the LSM RUM-tree.  Section~\ref{sec:concurrency} discusses how we support concurrent operations in the UM.  Section~\ref{sec:performance} presents  extensive experiments of the LSM RUM-tree in comparison to previous approaches. Section~\ref{sec:conclusion} contains concluding remarks.

\section{Background and Related Work}\label{sec:bg}
    In this section, we present 
    %the details 
    an overview 
    of the LSM-tree~\cite{o1996log}
    %, and summarize 
    along with
    two prior approaches to handle spatial workloads in the LSM R-tree. 
    Because the goal of this paper is to improve performance of the LSM R-tree via the use of the Update Memo, it is important to understand how the LSM R-tree works for each operation and how the Update Memo can accelerate the update procedure for the R-tree.
 
    \subsection{Log-Structured Merge Tree}
    \begin{sloppypar}
      The LSM-tree~\cite{o1996log} has been introduced to optimize write-intensive workloads by buffering data ingestion in the memory layer.
    Once the size of the data in the memory layer exceeds a certain threshold, the LSM-tree flushes the memory data to the disk layer using sequential I/O. As the flushed components accumulate in the disk layer, they get merged together based on a merge policy so that the LSM-tree maintains a limited number of disk components. Because of the LSM-tree's buffering ability to handle write-intensive workloads efficiently, it has been widely used in a variety of systems, e.g., AsterixDB~\cite{alsubaiee2014storage, luo2019efficient},
    %BigTable~\cite{chang2008bigtable},
    Cassandra~\cite{cassandra}, 
    %Dynamo~\cite{decandia2007dynamo},
    %HBase~\cite{HBase},
    LevelDB~\cite{LevelDB}, and RocksDB~\cite{RocksDB}. 
    \end{sloppypar}

    For clarity, we use the following terms throughout the paper as stated in the previous studies.
    \textit{Primary index} is a B$^+$-tree, where the key is the primary key (e.g., $O_{id}$) and the value is the data describing an object. 
    \textit{Primary key index} and \textit{Deleted-key B$^+$-tree} are also B$^+$-trees, but they hold the primary key only, not the entire object. The LSM secondary index (e.g., the LSM R-tree) has a secondary key (e.g., location) and a value that is the primary key of the object. Notice that for the same secondary key,  there can be multiple objects (e.g., different objects could be placed in the same location). 
    
    \subsection{The LSM R-tree}\label{sec:oldstrategies}
    \begin{sloppypar}
    Traditional disk-based spatial indexes, e.g., the  R-tree~\cite{guttman1984r}, adopt an ``in-place update'' policy. This results in many random I/Os. Given the high update rates of moving objects in space in the spatial data domain and the need to index these moving objects, it is not a surprise to apply LSM to the R-tree to deal with these update-intensive spatial workloads. 
    The LSM R-tree~\cite{alsubaiee2014storage} is a spatial 
    index
    to handle location data efficiently, and benefits from the LSM mechanism. By applying a generic framework for secondary index {\em LSMification}~\cite{alsubaiee2014storage}, the LSM R-tree is an optimized secondary index for write-intensive spatial data workloads. To handle frequent updates in the  LSM R-tree index, two strategies, termed Eager~\cite{alsubaiee2014storage} and Validation~\cite{luo2019efficient}, have been proposed. 
    The focus of this paper is to optimize the performance of LSM R-tree. Traditionally, updating a secondary key value, e.g., updating the  location of an object, requires maintenance of both an LSM \textit{primary}  and \textit{secondary} indexes to ensure consistency among both indexes.
    In the primary index side, the object ids (key) need to keep the new object location, while in the secondary index side, the new location (key) of an object should point to this object's id.
    Maintaining the primary index is straightforward. It 
    %just 
    buffers updates into the component in the memory layer. However, the maintenance of the LSM secondary index is complicated because there could be multiple objects with the same secondary key and/or multiple versions (i.e., the most recent or obsolete ones) for the same object. Moreover, the previous studies~\cite{alsubaiee2014storage, luo2019efficient} require point lookup on the primary index during the maintenance of the secondary index.  
    Thus, to simplify the presentation of the secondary index, we only highlight the maintenance of the LSM R-tree. 
\end{sloppypar}
    
    \definecolor{mypurple}{RGB}{112,48,160}
    \definecolor{myred}{RGB}{192,0,0}
    \definecolor{mygreen}{RGB}{0,176,80}
    
    \begin{figure*}[t]
        \centering
        \subfloat[LSM R-tree under eager strategy]
        {           
            \includegraphics[width=0.32\linewidth]{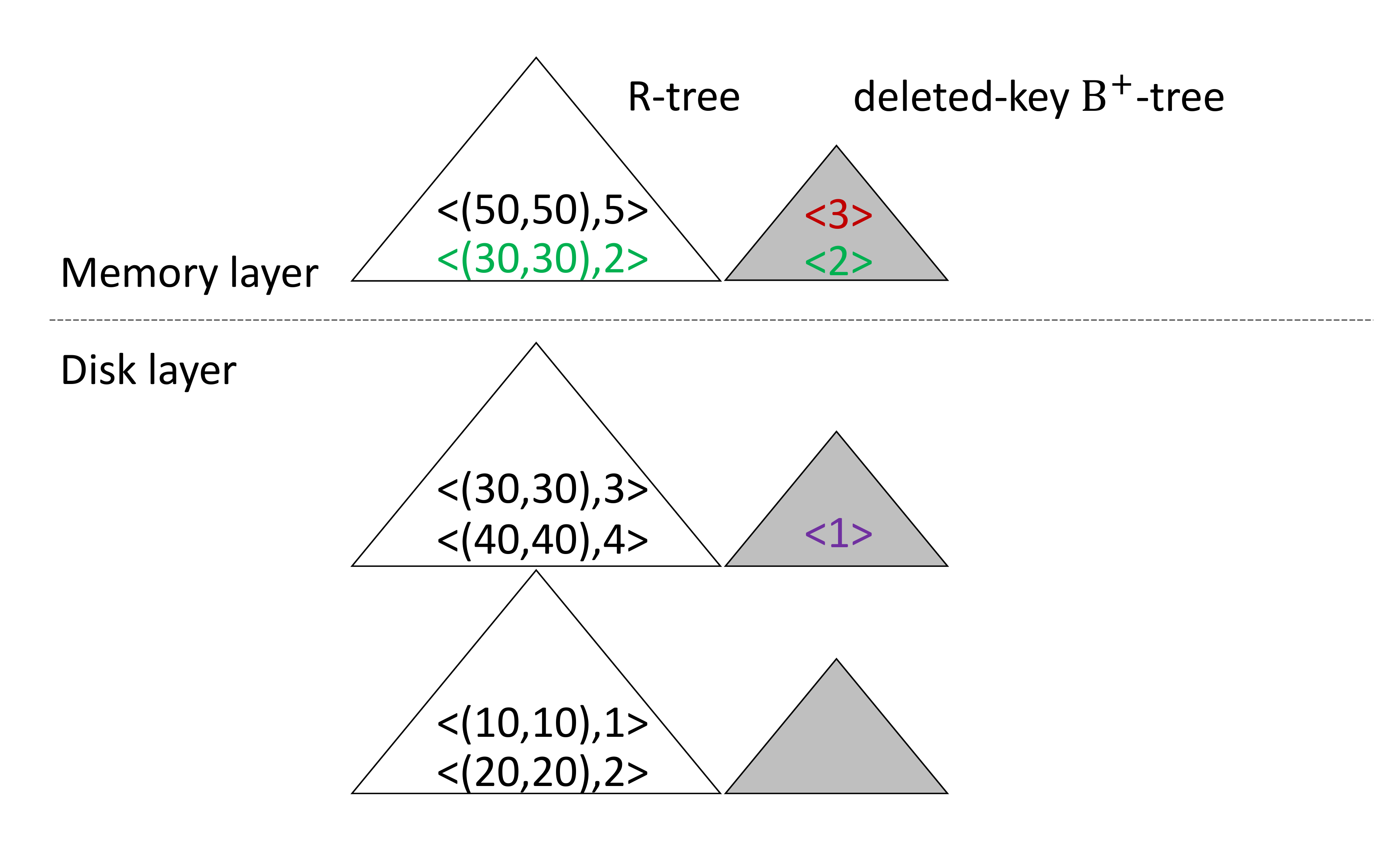} 
            \label{fig:eageroverview}          
        }
        \subfloat[LSM R-tree under validation strategy]
        % . Primary key index is used to validate R-tree objects by comparing with timestamp.]
        {       
            \includegraphics[width=0.32\linewidth]{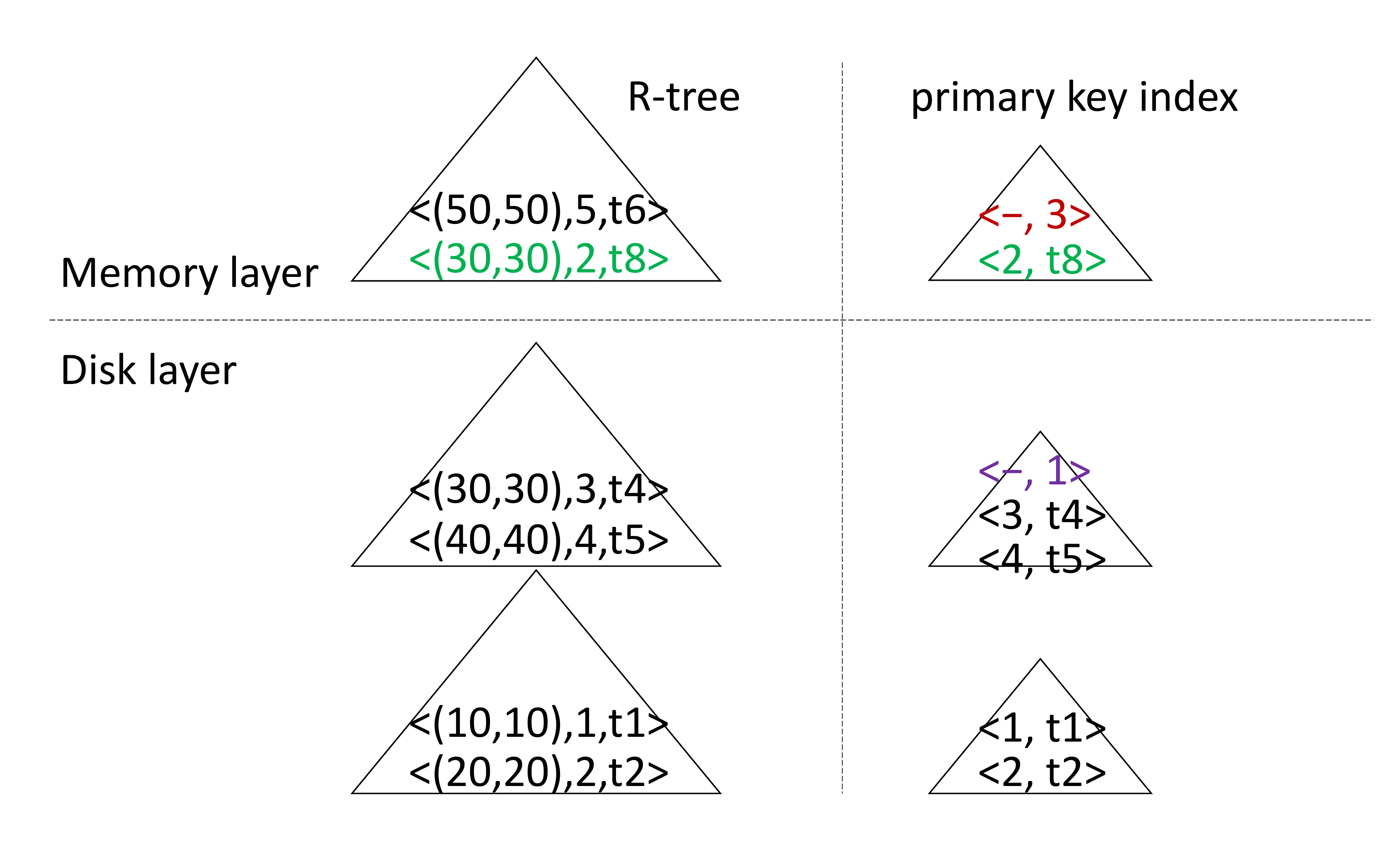} 
            \label{fig:validationoverview}          
        }
        \subfloat[Objects and their operations]
        {
            % \begin{table}
            \begin{adjustbox}{width=0.25\linewidth}
            \begin{tabular}{ | c | c | c |}
            \hline
                ts & oper. & object  \\
            \hline
                t1 & INSERT & $\langle (10,10),1\rangle $ \\ 
                t2 & INSERT & $\langle (20,20),2\rangle $ \\  
                t3 & \textcolor{mypurple}{DELETE} & \textcolor{mypurple}{$\langle 1\rangle $} \\ 
                t4 & INSERT & $\langle (30,30),3\rangle $ \\ 
                t5 & INSERT & $\langle (40,40),4\rangle $ \\
                t6 & INSERT & $\langle (50,50),5\rangle $ \\
                t7 & \textcolor{myred}{DELETE} & \textcolor{myred}{$\langle 3\rangle $} \\
                t8 &\textcolor{mygreen}{UPDATE} & \textcolor{mygreen}{$\langle (30,30),2\rangle $} \\
            \hline
            \end{tabular}
            \end{adjustbox}
            % \end{table}
            \label{table:runningorder}
        }
        \caption{Running examples}
    \end{figure*}
    
    \begin{sloppypar}
    \subsubsection{Eager Strategy for LSM Secondary Indexes}~\label{sec:bgEager}
    Alsubaiee et al.~\cite{alsubaiee2014storage} introduce to LSM an additional deleted-key B$^+$-tree for each R-tree to make the LSM R-tree consistent. The deleted-key B$^+$ tree stores the primary keys (the object ids) of the deleted/updated objects 
    to validate the state of an object given a query. The in-memory R-tree and its corresponding deleted-key B$^+$-tree are tightly coupled, and they are flushed to disk together as a component. Although the deleted-key B$^+$-tree buffers the delete operations in the memory layer, the extra maintenance cost and disk I/Os during a search operation degrades the overall search performance.

    Assume that we have an 
    object, say
    $o = \langle Loc, O_{id}, ... \rangle$, where $Loc$ indicates the 
    location of $o$, and $O_{id}$ is $o$'s object identifier. Because the R-tree indexes  locations, $Loc$ is the secondary key for the R-tree index while $O_{id}$ is a foreign key that points to the primary key of the object in the primary index, 
    where the latter may contain other attributes that describe $o$. To insert $o$ into the LSM R-tree, we add $o$ into the in-memory R-tree, and do not need to access the deleted-key B$^+$-tree.
\end{sloppypar} 
  
    To delete an entry $o=\langle Loc, O_{id} \rangle$ from the LSM R-tree, we perform the following steps: (1)~Remove $o = \langle Loc, O_{id} \rangle$ from the in-memory R-tree index, if it exists, and (2)~Invalidate the outdated $o$s in the disk components by adding $O_{id}$ into the deleted-key B$^+$-tree in memory. Step~1 is necessary in the {\em Eager} strategy for the consistency of the LSM R-tree. For example, given a sequence of updates for an object, without performing Step~1 in the memory component, i.e., insert: $\langle Loc_1, O_i\rangle $, delete: $O_i$, and insert: $\langle Loc_2, O_i\rangle $, there is no way to identify which of the entries $\langle Loc_1, O_i\rangle $ and $\langle Loc_2, O_i\rangle $ is the  recent location. As discussed in~\cite{alsubaiee2014storage}, entries in the deleted-key B$^+$-tree help validate objects in the disk components.

    To update a spatial object $o$ inside the LSM R-tree, e.g., due to $O_{id}$'s 
    changing its location from $Loc_{old}$ to $Loc_{new}$,
    we perform the following: Delete the old object $o = \langle Loc_{old}, O_{id} \rangle$ (by following the delete procedure above), and then insert the new object $o= \langle Loc_{new}, O_{id} \rangle$ into the LSM R-tree 
    (by following the insert procedure above). 
 
    \begin{sloppypar}
    To search the LSM R-tree (e.g., find all objects around $Loc = (2,2)$), all the R-trees, whether in memory or in disk, are searched to find candidate query results because all the R-trees could have qualifying objects. Note that the secondary index, i.e., the R-tree, could have multiple records for a given secondary key, i.e., $Loc$, so that the results for searching with the secondary key could be in any component. Then, the $O_{id}$ of each candidate will need to be searched against {\em all} the existing deleted-key B$^+$-trees, and will be reported as output only if $O_{id}$ does not exist in any of the deleted-key B$^+$-trees of a newer component than the candidate's component. Notice that due to flushing the in-memory deleted-key B$^+$-tree into disk along with its corresponding R-tree, there are multiple deleted-key B$^+$-trees, and we start a new one in memory along with its corresponding new in-memory R-tree.
     \end{sloppypar}   
    
    \noindent
    {\bf Running Example}:
    Refer to Table~\ref{table:runningorder} and Figure~\ref{fig:eageroverview} for illustration. Assume that insert and delete events arrive in the order as indicated in the table, where each object is denoted by $\langle Loc, O_{id}\rangle $. Figure~\ref{fig:eageroverview} illustrates how the {\em Eager} strategy handles inserts, deletes, and updates in the LSM R-tree. Each insert operation adds an object into the R-tree. For simplicity of illustration, 
    let the R-tree be of
    Capacity 2, i.e., 
    when an R-tree has two objects, it will be flushed to disk. 
    By Time~t3, Objects~1 and~2 will have been already inserted into the in-memory R-tree, and because its capacity is reached, the in-memory R-tree will be flushed to disk along with its corresponding empty deleted-key B$^+$-tree (Both appearing in the figure in the bottom row). At Time~t3, DELETE $\langle 1\rangle $ makes the existing object $\langle (10,10),1\rangle $ obsolete. Thus, $O_{id}$ $\langle 1\rangle $ is inserted into the deleted-key B$^+$-tree (This deleted-key B$^+$-tree appears in Row 2 of the figure). Therefore, when searching the R-tree at any future time and returning object $\langle (10,10),1\rangle $, this output will be invalidated by the key $\langle 1\rangle $ that exists in the deleted-key B$^+$-tree. 
    Observe that the {\em Eager} strategy 
    requires extra overhead to maintain  multiple deleted-key B$^+$-trees. 
    
    \subsubsection{Validation Strategy for LSM Secondary Index}~\label{sec:bgValidation}
    The Validation strategy~\cite{luo2019efficient} 
    addresses
    the update overhead 
    of the Eager strategy.
    It avoids using the deleted-key B$^+$-tree and uses the \textit{primary key index} to validate the query results. 
    The validation strategy 
    adds
    a timestamp in each object in both the \textit{primary key} and {\em secondary} indexes. 
    Figure~\ref{fig:validationoverview} illustrates how the validation strategy handles data operations from the same running example above. 
    For inserts, it inserts
    an object as $\langle Loc, O_{id}, ts\rangle $ into the R-tree where $ts$ is a timestamp for the insert using local wall-clock time.
    In addition, 
    for inserts, 
    the \textit{primary key index} stores an object as $\langle O_{id}, ts\rangle $.
    For the delete 
    at 
    Time 
    t3, the Validation strategy 
    only 
    inserts a control entry into the primary key index (e.g., $\langle -,1\rangle $, 
    to indicate that Object 1 is deleted). 
    Thus, 
    it simplifies the delete or update procedures over the {\em Eager} strategy.  
    However, the query performance of the {\em Validation} strategy still suffers due to 
    the needed validation steps using the primary index for direct validation or primary key index for timestamp validation.

    To summarize, 
    the {\em Eager} and {\em Validation} strategies 
    have inefficient update and search performances.
    For updates, the {\em Eager} strategy 
    removes the old object entry by traversing the in-memory R-tree and inserting a control entry into the deleted-key B$^+$-tree,
    and then 
    inserts
    the new object. 
    For this, the {em Eager} strategy 
    has degraded update performance. 
    The {\em Validation} strategy does not modify the R-tree upon 
    update, 
    but rather inserts the new object into the \textit{primary key index} and the R-tree with a timestamp. 
    In contrast to the {\em Eager} strategy, updates are faster in the {\em Validation} strategy.
    However, 
    search in 
    the {\em Validation} strategy degrades because the secondary index has false-positive results, and 
    needs 
    to validate the results by comparing against the primary key index.

    \subsection{The RUM-Tree: The R-tree with Update Memo}\label{sec:bgRUM}
       \sloppy The RUM-tree~\cite{xiong2006r, silva2009rum} is another approach to handle update-intensive spatial workloads. It augments an R-tree with an in-memory Update Memo (UM) structure. The RUM-tree is a disk-based R-tree index that has a UM in memory. It maintains a global timestamp counter, and marks each object to show a temporal relationship 
       %between 
       among the
       objects. Each UM entry 
       %consists of 
       is of the form: $\langle O_{id}, ts, cnt\rangle $, where $ts$ is the 
       %most-recent timestamp of an object version that has the same $O_{id}$ 
       time\-stamp of the most-recent update to $O_{id}$ in the RUM-tree,
       and $cnt$ is the number of obsolete versions of Object $O_{id}$ in the index. By handling the insert and update operations in UM, the RUM-tree achieves significantly lower update cost without having big penalty 
       %in search performance. 
       during search.
       The UM has several cleaning strategies for removing the obsolete entries from both the R-Tree and the UM to restrict the latter's size. As a result, the RUM-tree shows improved performance on updates over the traditional R-tree~\cite{xiong2006r, silva2009rum, chen2008benchmark}. However, it is not possible to directly use the RUM-tree as is in an LSM environment because the RUM-tree's UM is tuned for use with the traditional R-tree, and if used as is, its UM maintenance procedures will result in a large UM size without proper modification of its mechanisms. In this paper, we expand on the idea of an Update Memo, and propose new strategies to harmonize it for use with an LSM-based R-tree secondary index. In the following sections, we illustrate how the UM is adapted for this purpose. We illustrate the LSM-based R-tree's insert, delete, update, and search algorithms, and demonstrate how they make use of UM while controlling its size, and hence can always fit in memory.

\section{The LSM RUM-TREE}\label{sec:lsmrum}
    We introduce the new LSM RUM-tree; an LSM R-tree augmented with an Update Memo. The goal of the LSM RUM-tree is to efficiently handle update-intensive spatial workloads, and improve search performance. 
From Section~\ref{sec:oldstrategies}, 
    existing 
    strategies 
    for 
    LSM secondary indexes 
    have degraded update and search performances.
    To address this issue, 
    in the LSM RUM-tree, we introduce an Update Memo within the LSM-based R-tree to simplify the processing of deletes and updates in the memory layer, and make disk-based search  cost-efficient. 
        
    \subsection{The Update Memo  Structure (UM)}\label{sec:updatememo}
\begin{figure}
        \centering
        \includegraphics[clip,  width={0.95\linewidth}]{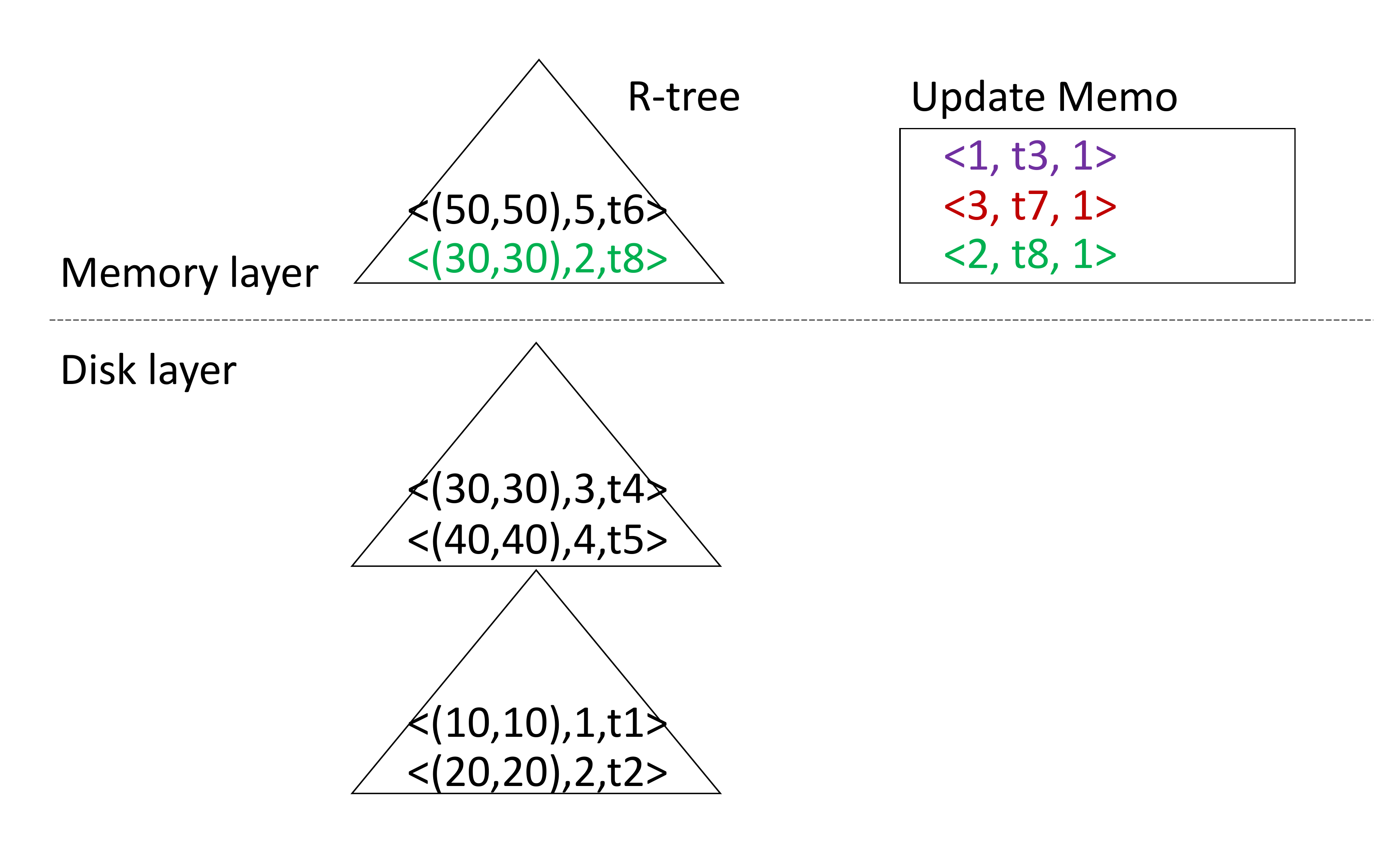}
        \caption{LSM RUM-tree Overview}
        % \vspace*{-1em}
    	\label{fig:umoverview}
    \end{figure}

    Refer to Figure~\ref{fig:umoverview} for illustration. In the LSM RUM-tree, the Update Memo (UM) is based on a hash map that resides in memory.
    The key to UM is an object identifier, and the value is a pair of recent timestamp and obsolete-objects counter. 
    The UM has several roles. 
    It keeps track of the deletes and updates ingested into the LSM RUM-tree. Because it resides only in the memory layer, the update to a UM entry takes constant time.
    Also, UM validates candidate objects resulting from a search operation. 
    As in Section~\ref{sec:oldstrategies}, previous strategies need to perform extra work to answer a query correctly. The {\em Validation} strategy needs to validate output candidates of a search by comparing the output candidates against \textit{the primary key index} in a way similar to that of the {\em Eager} strategy. 
    Because UM is light-weight and resides entirely in memory, we expect enhancement in search performance over the previous approaches.  

    From Figure~\ref{fig:umoverview}, each object in the LSM RUM-tree is represented as a triplet $\langle Loc, O_{id}, ts\rangle $, where $Loc$ is the object's location, $O_{id}$ is the object identifier, and $ts$ is a global timestamp counter -- an integer value incremented by 1 for each insert, delete, or update operation. The higher the value of $ts$, the fresher the object in the index. An entry $e$ in UM is represented by a triplet $\langle O_{id}, ts, cnt\rangle $, where $cnt$ is a counter of the number of outdated objects with the same $O_{id}$ in the LSM RUM-tree. By default, the entry $e=\langle O_{id}, ts, \textbf{1}\rangle $ is first inserted into UM when deleting or updating an object with no existing entry in UM. Otherwise, if  $O_{id}$ exists in UM in  an entry, say $e$, from a previous delete or update, we increment $e.cnt$ by 1. When an obsolete object is found and is removed from the LSM RUM-tree, we decrement the object's $e.cnt$ in UM by 1. Entry $e$ is removed from UM when $e.cnt = 0$, i.e., there are no obsolete entries for $O_{id}$ currently in the LSM RUM-tree. 
    
    \subsection{Lazy Maintenance in UM}\label{sec:dataoperations}
    UM buffers deletes and updates in memory. 
    Below, we discuss how the  delete, update, and search operations in the LSM RUM-tree utilize UM. 
% \vspace*{-1em}

        \subsubsection{Insert}
        
        \begin{algorithm}[h]
            \SetKwInOut{Input}{input}
            \Input{$Loc$: Location (secondary key) of the object \newline 
                    $O_{id}$: Object id (primary key)}
            
            $ts \gets $ timestamp counter++\;
            $O_{new} = \langle Loc, O_{id}, ts\rangle $\;
            $insertLSMRtree(O_{new})$\;
            
            \caption{Insert operation}
            \label{alg:insert}
        \end{algorithm}
% \vspace*{-1em}
        Refer to Algorithm~\ref{alg:insert}. During an insert, first we increment the global time\-stamp counter by 1 (Line~1). Note that the timestamp counter is an integer that gets incremented with each insert, delete, or update operation. Once we read the global timestamp counter, say $ts$, a new object $o_{new}=\langle Loc, O_{id}, ts\rangle $ is added to the in-memory R-tree (Lines~2-3). UM tracks only object deletes and updates, and 
        performs no special actions for inserts. The goal 
        of 
        the LSM RUM-tree is to support both update- and insert-intensive spatial workloads. If UM 
        treats
        inserts as 
        in the case of
        the original RUM-tree~\cite{xiong2006r,silva2009rum}, the UM size grows linearly with the number of objects for insert-intensive workloads. This would 
        result
        in significant memory-space overhead. In the LSM RUM-tree, avoiding to maintain UM upon object inserts prevents the unnecessary growth in UM's size. In Section~\ref{sec:queryoperation}, we discuss in detail how we validate an object that has no entry in UM.
% \vspace*{-1em}

        \subsubsection{Delete}
        \begin{algorithm}
            \SetKwInOut{Input}{input}
            \Input{$O_{id}$: Object id (primary key)}
            
            $ts \gets $ timestamp counter++\;
            \If{entry $e$ for $O_{id}$ exist in $UM$}{
                $e.ts \gets $ $ts$\;
                $e.cnt$++\;
            }
            \Else(){
                put $e_{new} = \langle O_{id}, ts, 1\rangle $ into $UM$\;
            }
            
            \caption{Delete operation}
            \label{alg:delete}
        \end{algorithm} 
% \vspace*{-1em}        

        In the LSM RUM-tree, we minimize the processing needed to validate results. To delete an object, we only add or modify the object's corresponding UM entry. As in Algorithm~\ref{alg:delete}, if there is a UM entry, say $e$, for a given $O_{id}$, we set Field $e.ts$ to the global timestamp (Line 3) and  increment $e.cnt$ by 1 (Line 4). If there is no such $e$ in $UM$, we insert $\langle O_{id}, ts, 1\rangle $ into $UM$. Note that the value of $e.cnt$ corresponds to the number of obsolete copies of a given $O_{id}$ in the R-trees. Naturally, each delete operation makes one additional obsolete object in the R-tree. By tracking $ts$ for freshness and $cnt$ for the number of obsolete objects of $O_{id}$, we not only have a clear sense of a given object copy in the index whether it is fresh or not, but also have a good grasp of the number of  obsolete copies of the object in the LSM RUM-tree. In the example in Figure~\ref{fig:umoverview}, ``t3 : DELETE $\langle 1\rangle $'' and ``t7 : DELETE $\langle 3\rangle $'' add $\langle 1, t3, 1\rangle $ and $\langle 3, t7, 1\rangle $ into the UM, respectively.
% \vspace*{-1em}

        % \subsubsection{Update}\label{sec:umupdate}
            
        % \begin{algorithm}
        %     \SetKwInOut{Input}{input}
        %     \Input{$Loc$: Location (secondary key) of the object \newline 
        %             $O_{id}$: Object id (primary key)}
            
        %     $ts \gets $ timestamp counter++\;
        %     $O_{new} = \langle Loc, O_{id}, ts\rangle $\;
        %     \If{entry $e$ for $O_{id}$ exist in $UM$}{
        %         $e.ts \gets $ $ts$\;
        %         $e.cnt$++\;
        %     }
        %     \Else(){
        %         put $e_{new} = \langle O_{id}, ts, 1\rangle $ into $UM$\;
        %     }
        %     $insertLSMRtree(O_{new})$\;
            
        %     \caption{Update operation}
        %     \label{alg:update}
        % \end{algorithm}  
        
% \vspace*{-1em}            

        \subsubsection{Update}\label{sec:umupdate}
        To process an update, we check whether or not UM contains an entry, say $e$, with the same $O_{id}$. If $e$ exists in $UM$, we update $e.ts$ to the current timestamp $ts$ and increment $e.cnt$ by 1. If $e$ does not exist, we add a new entry $e_{new}=\langle O_{id}, ts, 1\rangle $ into UM. Then, we add the new object entry $\langle Loc, O_{id}, ts\rangle $ into the in-memory R-tree. Note that update consists of both delete and insert. Thus, we follow Lines~1-6 of Algorithm~\ref{alg:delete} and Lines~2-3 of Algorithm~\ref{alg:insert}.
        In the example in Figure~\ref{fig:umoverview}, ``t8 : UPDATE $\langle (30,30),2\rangle $'' adds the entry $\langle 2, t8, 1\rangle $ into the UM and inserts the new object $\langle (30,30), 2, t8\rangle $ into the in-memory R-tree. Notice that if there is another UPDATE for the same object at t9, we update the UM entry from $\langle 2, t8, 1\rangle $ to $\langle 2, t9, 2\rangle $ and insert the new object with the timestamp $t9$ into the R-tree.

    \subsection{Search}\label{sec:queryoperation}
% \vspace*{-1em}    
    \begin{algorithm}
        \SetKwInOut{Input}{input}
        \SetKwInOut{Output}{output}
        \Input{$candidates$: The list of candidates for a search}
        \Output{$results$: The list of results of a search}
        
        \For{$O_{cand} \gets$ $candidates$}{
            $O_{id} \gets O_{cand}.O_{id}$\;
            \If{entry $e$ for ${O_{id}}$ exist in $UM$}{
                \If{$O_{cand}.ts == e.ts$}{
                    $results$.insert($O_{cand}$)\;
                }
            }
            \Else(){
                $results$.insert($O_{cand}$)\;
            }
        }
        \Return $results$
        \caption{Validation with Update Memo}
        \label{alg:query}
    \end{algorithm}  
% \vspace*{-1em}  
    We explain how the LSM RUM-tree performs a search operation. Both the {\em Eager} and {\em Validation} strategies take extra disk I/Os to validate search results due to accessing the additional structures, either the deleted-key B$^+$-tree or the primary key index (B$^+$-tree). The UM is memory-resident (due to the cleaning strategies in Section~\ref{sec:cleaning}). Thus, the LSM RUM-tree does not require a disk I/O except for accessing the disk-based LSM R-trees. By removing the additional tree structures from the LSM RUM-tree, we save on extra tree traversal times as well as on maintenance costs and expect to perform better during search.
        
    Algorithm~\ref{alg:query} illustrates how to utilize the UM to validate search results that are returned from the LSM RUM-tree. First, we check whether a candidate $O_{cand}$ from the R-tree search is fresh or not. If UM does not contain an entry with the same $O_{id}$, then $O_{cand}$ is fresh, and is part of the search results. If there is an entry, say $e$, with the same $O_{id}$, then the $ts$ field of the candidate is compared with $e.ts$ from UM. If $O_{cand}.ts < e.ts$, then the candidate is obsolete, and is discarded, otherwise, is discarded.
    Observe that there is no case where $O_{cand}.ts > e.ts$ because we always maintain a UM entry to reflect the most recent timestamp of an object. Also, if there is a fresh object (e.g., ``t5 : INSERT $\langle (40,40),4\rangle $'' in Table~\ref{table:runningorder}) and there is no other updates on the same object, there is no UM entry with the same $O_{id}=4$. 
    For example, in Figure~\ref{fig:umoverview}, if there is a range search from (30,30) to (40,40), the candidates in the LSM RUM-tree are [$\langle (30,30),3,t4\rangle $, $\langle (40,40),4,t5\rangle $, $\langle (30,30),2,t8\rangle $]. The candidate $\langle (30,30),3,t4\rangle $ has smaller timestamp than its entry in UM ($\langle 3,t7,1\rangle $) because of the DELETE $\langle 3\rangle $ at t7. Thus, this candidate is discarded. For $\langle (40,40),4,t5\rangle $, UM does not contain an entry with $o_{id}=4$, i.e., this object is fresh, and the candidate is a part of the search results. For Object  $\langle (30,30),2,t8\rangle $, its entry $\langle 2,t8,1\rangle $ shows the object is a part of the results as well. Thus, the final search results are [$\langle (40,40),4,t5\rangle $, $\langle (30,30),2,t8\rangle $]. 
            
    Because UM is based on a hash map, massive amounts of delete/update operations will increase the size of UM. Also, as the size of the hash map increases, its lookup performance deteriorates due to the large number of entries in each hash map bucket and these result in degrading the search performance. For these reasons, in the next section, we introduce LSM-aware UM cleaning strategies to bound the size of UM and improve search performance.
        
\section{LSM-Aware UM Cleaning Strategies}\label{sec:cleaning}
To ensure that UM fits in memory, we control its size using cleaning strategies. We introduce various cleaning strategies that reduce the size of the UM efficiently without much overhead. As in Section~\ref{sec:updatememo}, an entry $e$ in the UM will be removed when the field $cnt$ hits 0. 
Thus, our focus is on how to decrease $e.cnt$ for each operation running on the LSM RUM-tree. 
We introduce four cleaning strategies for the LSM RUM-tree: (1)~Buffered Cleaning, (2)~Vacuum Cleaning, (3)~Clean Upon Flush, and (4)~Clean Upon Merge. The first two are for UM cleaning through the in-memory R-tree. In contrast, the remaining two are for UM cleaning through the disk-side R-trees. Although the RUM-tree~\cite{silva2009rum} also employs cleaning strategies to clean the obsolete objects from the underlying R-tree, it is not straightforward
to apply the same strategies directly into the LSM-based R-tree due to the added complexity of the LSM-based R-tree.

\subsection{Buffered Cleaning}\label{sec:bufferedCleaning}
    When an application uses the LSM RUM-tree as a secondary index to handle continuous update-intensive workloads (e.g., tracking moving objects continuously), there is a high chance that a node of the in-memory R-tree has multiple obsolete objects. To clean UM and those in-memory R-tree nodes, we introduce the Buffered Cleaning strategy that cleans an in-memory R-tree node based on the accumulated updates inside this node. This is particularly applicable for hot-spot cleaning. We maintain an \textit{update counter} for each node of the in-memory R-tree. When the LSM RUM-tree buffers an update operation, the update counter on the node is incremented by 1. Once the counter hits some threshold, we remove the obsolete objects from the node and clean the UM entry by decrementing  its $cnt$ value by 1 for each of the removed obsolete objects. When a new R-tree node is created (e.g., due to a node split) or when Buffered Cleaning cleans a specific node, we set the node's update counter to 0. 
        
        \begin{figure}
            % \centering
            \subfloat[R-tree nodes changes]
            {           
                \includegraphics[clip,trim={0 {.15\linewidth} 0 0}, width=0.95\columnwidth]{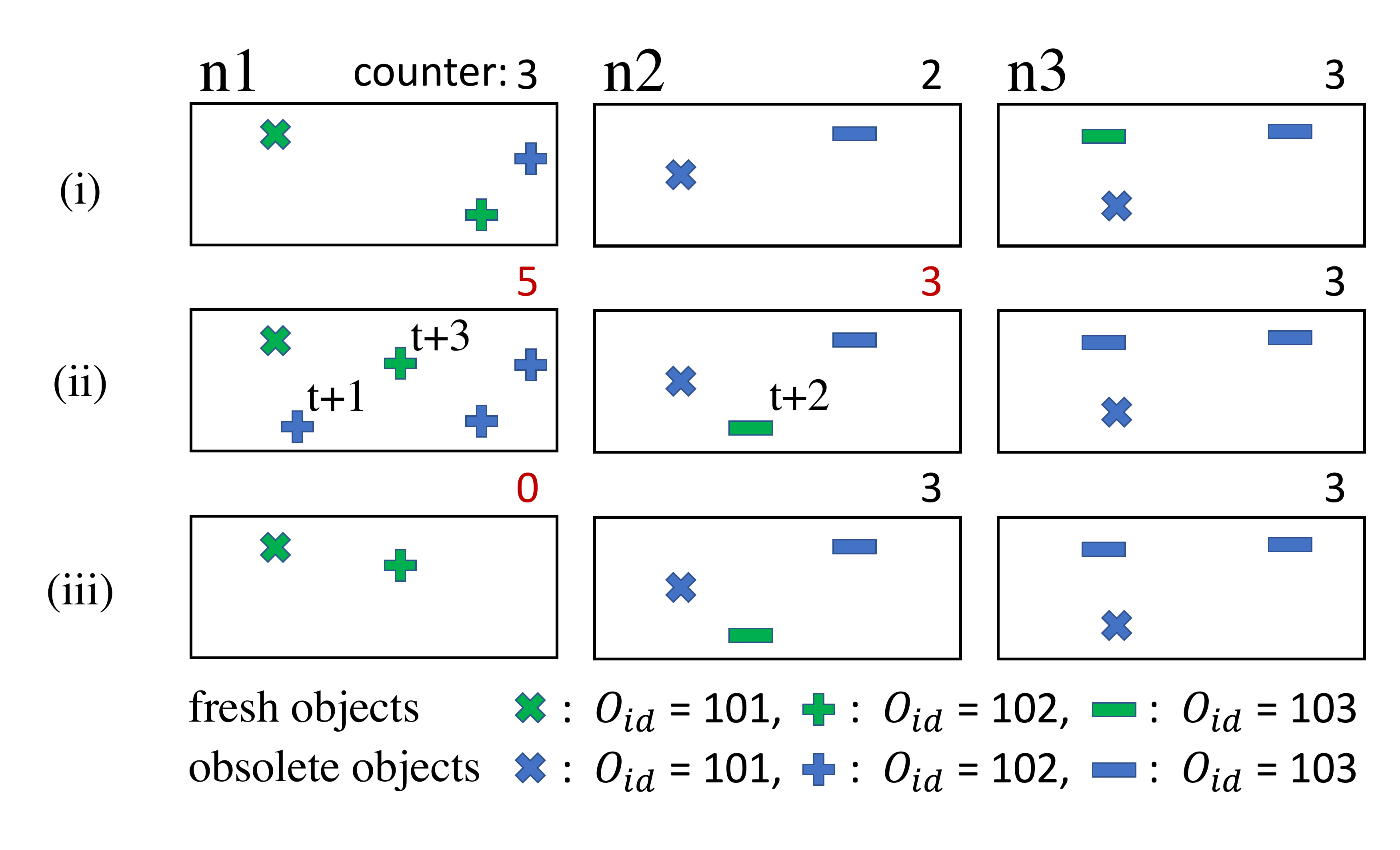} 
                \label{fig:bcNodes}          
            }
            
            \subfloat[Update Memo changes]
            {       
                \includegraphics[clip,trim={0 {.77\linewidth} 0 {.6\linewidth}}, width=0.95\columnwidth]{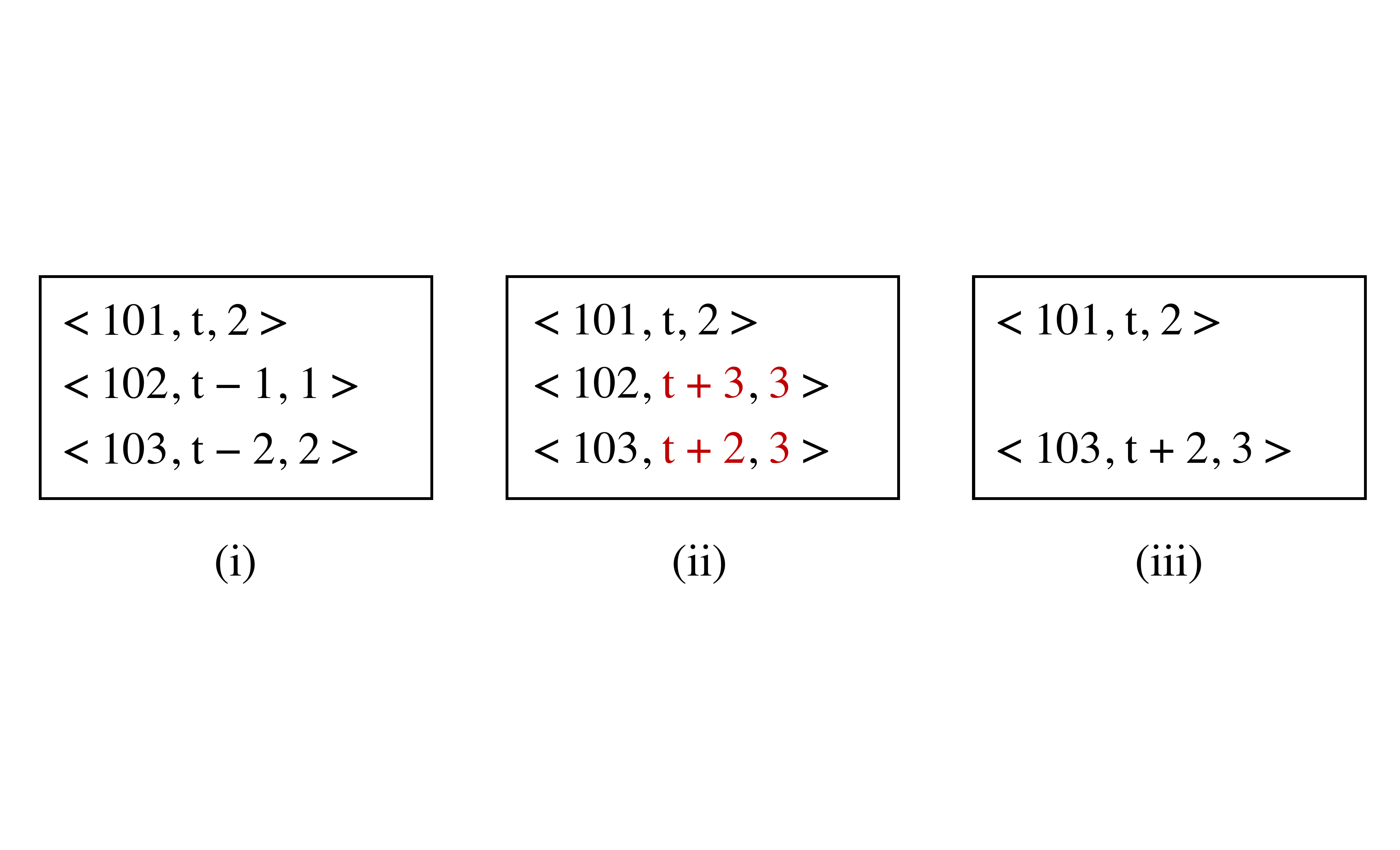} 
                \label{fig:bcUM}          
            }
            \caption{Examples of Buffered Cleaning}
            % \vspace*{-1em}
        \end{figure}
    
        % \begin{figure}[h]
        %     \centering
        %     % \includegraphics[clip, trim={.4\linewidth} {.4\linewidth} {.4\linewidth} {.1\linewidth}, width={\linewidth}]{pic/03gridquadtree}
        %     \includegraphics[clip,  width={\linewidth}]{pics/bufferedcleaning.pdf}
        %     % \vspace*{-1cm}
        %     \caption{Examples of Buffered Cleaning in R-tree nodes}
        %     % \vspace*{-1em}
        % 	\label{fig:bcNodes}
        % \end{figure}
        % \begin{figure}[h]
        %     \centering
        %     \includegraphics[clip, trim={0 {.77\linewidth} 0 {.78\linewidth}}, width={\linewidth}]{pics/bufferedcleaningUM.pdf}
        %     % \includegraphics[clip,  width={\linewidth}]{pics/bufferedcleaningUM.pdf}
        %     % \vspace*{-0.7cm}
        %     \caption{Examples of Buffered Cleaning in Update Memo}
        %     % \vspace*{-1em}
        % 	\label{fig:bcUM}
        % % 	\vspace*{-1em}
        % \end{figure}

    For example, Figures~\ref{fig:bcNodes} and~\ref{fig:bcUM} illustrate how Buffered Cleaning works in the LSM RUM-tree. There are three distinct objects (with $O_{id}$s 101, 102, and 103) over three R-tree nodes ($n1$, $n2$, and $n3$). We color the fresh objects green and the obsolete ones blue.  Assume that the threshold for the update counter is 5. Figures~\ref{fig:bcNodes}(i) and~\ref{fig:bcUM}(i) give the state of the R-tree nodes and that of UM at some point $t$. Notice that the counters for Nodes $n1$, $n2$, and $n3$ are 3, 2, and 3, respectively. Three more updates at $t+1$, $t+2$, and $t+3$ take place, and this results in Figures~\ref{fig:bcNodes}(ii) and~\ref{fig:bcUM}(ii), as  in Section~\ref{sec:umupdate}. Because $n1$'s counter hits the threshold, we clean $n1$ as well as the UM. The threshold for the update counter is a variable that decides the frequency of invoking the Buffered Cleaning strategy. The lower the threshold is set, the more frequent the Buffered Cleaning strategy is invoked. Notice that frequent Buffered Cleaning results in extra cost to clean a node and the UM. Thus, we need to tune the threshold depending on the characteristics of the workload. For example, if the incoming location updates are continuous and gradual (e.g., due to movements of objects in small displacements), the lower threshold works efficiently. If location updates are periodic and geographically random (e.g., social media), the higher threshold can avoid extra cleaning-overhead.
% \vspace*{-1em}

        \begin{algorithm}[h]
            \SetKwInOut{Input}{input}
            \SetKwInOut{Output}{output}
            \Input{$objects$: the list of objects in a node}
            
            \For{$O \gets$ $objects$}{
                $O_{id} \gets O.O_{id}$\;
                \If{entry $e$ for $O_{id}$ exist in $UM$}{
                    \If{$O.ts < e.ts$}{
                        remove $O$ from the node\;
                        $e.cnt--$\;
                        \If{$e.cnt == 0$}{
                            remove $e$ from the Update Memo\;
                        }
                    }
                }
            }
            
            \caption{A node and the Update Memo cleaning}
            \label{alg:nodecleaning}
        \end{algorithm}  
% \vspace*{-1em}
        
Algorithm~\ref{alg:nodecleaning} illustrates how we clean an R-tree node and its corresponding entries in the UM. While iterating over objects in a node, we discard the obsolete objects from the node by comparing the object's timestamp with the timestamp in the object's entry in the UM (Lines~2-5). If an obsolete object is removed, we decrement by 1 the object's corresponding $e.cnt$ entry in UM to track the number of obsolete objects with the same $O_{id}$ (Line~6). If $e.cnt=0$,  there are no obsolete objects having the same $O_{id}$ in the LSM RUM-tree. Thus, we remove $e$ from the UM. Note that there are two cases that have $e.cnt=0$: (1)~There is only one copy of the object left having $O_{id}$ in the LSM RUM-tree and the object is fresh, or (2)~There is no such object because all obsolete objects are removed, and the last operation for the object with the same $O_{id}$ is a delete operation. The Buffered Cleaning strategy handles both cases consistently. By the end of the cleaning steps, Buffered Cleaning sets the update counter to 0.

\subsection{Vacuum Cleaning}\label{sec:vacuumCleaning}
Vacuum Cleaning complements Buffered Cleaning because it targets mostly the hot-spot nodes in the LSM RUM-tree. There are still some cases that Buffered Cleaning cannot handle very well: (1)~A node is on a cold-spot so the counter for the Buffered Cleaning does not hit the threshold or (2)~Objects in a node have been obsoleted because of updates in other nodes. Not cleaning these cold-spot nodes can result in growing the size of UM and being not able to control it. To make up for the  cold-spot nodes not handled by Buffered Cleaning, we introduce Vacuum Cleaning for fair cleaning of in-memory R-tree nodes 
to bound UM's size. 

In Vacuum Cleaning, we maintain a global counter and a vacuum cleaner. The global counter stores the number of updates in the entire LSM RUM-tree, and the vacuum cleaner holds the next leaf node to be cleaned in the in-memory R-tree. Once the global counter hits some threshold by update operations, we clean the next node by the vacuum cleaner. The node and UM cleaning are the same as the ones in Algorithm~\ref{alg:nodecleaning}. After this node's cleaning is finished, we reset the global counter to 0 and set the vacuum cleaner to the next leaf node. We can skip the nodes that have been recently cleaned by the Buffered Cleaning Strategy. We can adjust the frequency of  Vacuum Cleaning by varying the threshold $th$ of the global counter. If the number of leaf nodes of the in-memory R-tree is $N$, it takes at most $N \times th$ updates to fully clean the R-tree. For example, small $th$ leads to a shorter \textit{full cycle} as in Figure~\ref{fig:cleaningcycle}. Depending on the capacity of the R-tree, tuning $th$ is needed for best performance results. Similar to Buffered Cleaning, a lower value of $th$ results in frequent Vacuum Cleaning executions, and may lead to excessive overhead. Depending on the size of the in-memory R-tree and its node capacity, a fine tune will be needed to clean the R-tree efficiently and avoid extra overheads.

 The Buffered and Vacuum Cleaning strategies have several advantages. They help reduce the UM size. Having obsolete objects in the in-memory R-tree leads to unnecessary UM entries. By cleaning the UM, we bound its size. Also, both cleaning strategies clean the nodes of the in-memory R-tree. Update-intensive workloads cause many obsolete objects in an R-tree node, and this leads to unnecessary R-tree node splits. By cleaning the nodes in the in-memory R-tree, the nodes remain fresh and avoid unnecessary splits due to being filled with obsolete objects. Also, by cleaning R-tree nodes, we expect to improve search performance because there should be less false-positive candidates (the obsolete entries in a leaf node) while searching the R-tree.
    
\subsection{Clean Upon Flush}\label{sec:cleanUponFlush}\label{sec:cleanUponFlush}
        \begin{figure}[hbt!]
                \centering
                \includegraphics[clip, trim={0 {.7\linewidth} 0 {.68\linewidth}}, width={\linewidth}]{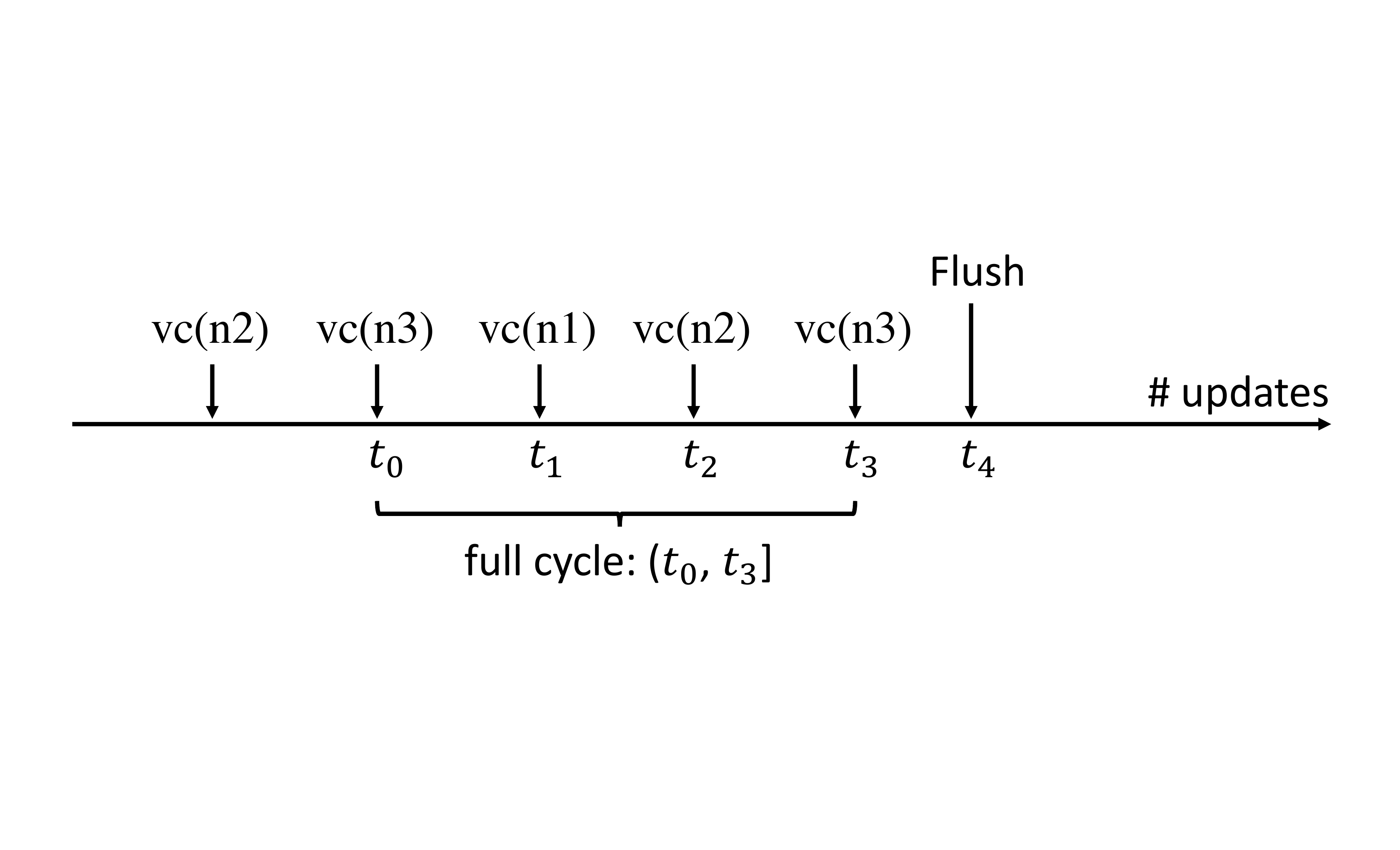}
        \caption{Cleaning cycle by Vacuum Cleaning}
        \label{fig:cleaningcycle}
        % \vspace*{-1em}
         \end{figure}
         
The Buffered and Vacuum Cleaning strategies are for cleaning the UM and the in-memory R-tree. For LSM-based indexes, there are unavoidable times when these two approaches miss obsolete objects, and thus the UM size increases, e.g., when LSM decides to dump the memory contents to disk while the in-memory R-tree is not fully cleaned. We introduce the Clean Upon Flush cleaning technique that will be executed during the flush operation of the LSM RUM-tree.
Assume that there are three nodes in the in-memory R-tree, that Vacuum Cleaning runs on the R-tree nodes in a round-robin manner, and that Buffered Cleaning is off. In the LSM RUM-tree with component-based flush, a component is flushed to disk when the size of the memory component exceeds its capacity. In Figure~\ref{fig:cleaningcycle}, periodic Vacuum Cleaning (i.e., vc(node)) takes place as updates accumulate in the LSM RUM-tree, and there is a flush operation at $t4$. Note that even though it has completed a full vacuum cleaning cycle at $t_3$, there should be obsolete objects at the flush time $t_4$. Precisely, at the end of the last vacuum cleaning cycle ($t_3$), all obsolete objects made before $t_1$ are cleaned. However, some obsolete objects (e.g., ones made between $t_1$ and $t_3$ in Node $n1$ are not cleaned at $t_4$ because the last clean on $n1$ was at $t1$). Also, all obsolete objects after $t3$ are not cleaned. Once the in-memory component is flushed into disk, it is expensive to modify the disk component because it will  lead to extra disk I/O, and will even violate the LSM policy: the disk components of the LSM-tree are immutable. 

Thus, we introduce the Clean Upon Flush cleaning strategy that cleans the obsolete objects before they are moved to disk. In Clean Upon Flush, we bring the flushed component up-to-date at the time of flushing without having too much computational overhead. Mainly, we add the R-tree and UM cleaning step 
just before flushing.
When the LSM RUM-tree is flushed into disk, it orders the objects by the Z-order or Hilbert curves~\cite{hilbert1891stetige, peano1890courbe} 
to be stored in disk efficiently. For Clean Upon Flush, when the flush operation orders the objects, we check whether the object is obsolete or not by comparing it with the entry in the UM as in Algorithm~\ref{alg:nodecleaning}. If the object is obsolete, we discard the object so it does not get flushed into disk, and we clean the corresponding UM entry. If the $cnt$ of the UM entry is 0, we remove the entry from UM. 

We expect that Clean Upon Flush will improve  performance in various ways. First, the UM size is reduced due to cleaning. Shrinking the UM size is a top priority in all cleaning strategies. Furthermore, all the objects in the flushed component are fresh at the time of the flushing. Thus, the size of the flushed component could be shrunk dramatically in some cases. For example, if most objects in the in-memory R-tree are obsolete and only a few objects are fresh, the size of the flushed R-tree will be proportional to the number of fresh objects. Consider an extreme case of having million updates from a hundred distinct objects.
Following a Clean Upon Flush, only hundred fresh objects will be presented in the flushed R-tree in contrast to million. Thus, searching the flushed LSM RUM-trees will require fewer disk 
I/Os due to the smaller flushed components, and  search performance would be improved significantly.

% \vspace*{-0.4em}        
\subsection{Clean Upon Merge}\label{sec:cleanUponMerge}
\label{sec:cleanUponM}        
Despite having in-memory cleaning strategies and cleaning at flush time, we can still have obsolete objects in disk. For example, if there is a fresh object $O_{disk}$ having some $O_{id}$ somewhere in the disk R-tree, any update to $O_{id}$ will make  $O_{disk}$ obsolete. 
This is likely to happen in moving-objects applications as the objects' locations are updated frequently. {\bf Thus, the objects in disk components become obsolete as new updates accumulate into the in-memory component}. 

We introduce the Clean Upon Merge cleaning strategy to clean the obsolete objects on disk as well as their corresponding UM entries. 
Direct modifications on the disk component are costly and not suitable for LSM. Thus, we leverage the LSM merge operation. The target of the LSM  merge operation is to bound the number of disk components via merging, and hence control the search time. Having many disk components degrades search performance as the search needs to investigate all disk-based R-trees. For the merge operation, multiple disk components are bulk-loaded and are merged into a single disk component.  With Clean Upon Merge, we have a  validation step as the one  in Algorithm~\ref{alg:nodecleaning} on the bulk-loading procedures of the merge operation. When bulk-loading the existing components, we check each object on-the-fly against the UM entries. If an object is obsolete, we discard it. Obviously, we update the UM entries accordingly in a way similar to Clean Upon Flush and the other cleaning strategies. 
        
Clean Upon Merge cleans the obsolete objects in the disk components with minor modifications to the LSM merge operation. Note that it is guaranteed that there will be no obsolete objects (i.e., an empty UM) in a system when there is only one disk R-tree after the merge with Clean Upon Merge. If combined with Clean Upon Flush, both can enhance search performance due to accessing fewer and more compact R-trees on disk.  Experimental results show that the query performance with Clean Upon Flush and Clean Upon Merge achieves over 400x speedup when compared to the validation strategy.

\section{Concurrency Control}\label{sec:concurrency}

% \jianguo{Need some motivation on the importance of studying concurrency control. Also, explain clear on what're the unique challenges about cc in our setting.} \jaewoo{edited}

The memory component of the LSM RUM-tree with a latch protocol protects a node in the tree that performs concurrent read/write operations in a multithreaded environment. When traversing the LSM RUM-tree, a thread is required to acquire/release latches on the nodes for a safe search or insert/update operations. For example, AsterixDB follows the latch protocols described in ARIES/KVL~\cite{mohan1989aries} and GiST~\cite{kornacker1997concurrency} for the B-tree and the R-tree, respectively.  Although a traditional LSM-tree protects its memory component by the latch protocols, introducing the Update Memo poses new challenges to support concurrent operations over UM because a corrupted data structure (e.g., incorrect $ts$ or $cnt$ of an UM entry) would lead to incorrect search results or UM cleaning failures. In this section, we demonstrate how to implement and achieve concurrency in UM.

% \jianguo{Not clear why you mention CC in LSM-tree first.}\jaewoo{merged with the motivation above}

% \jianguo{Why do we need "additiona protection"? Need to be clear.}\jaewoo{edited and moved to the motivation above}

\subsection{Concurrency Control in Update Memo}\label{sec:ccum}
We present how the Update Memo can support concurrent operations in a multithreaded environment. As discussed in Section~\ref{sec:updatememo}, UM is based on a hash map structure where the key to UM is an object identifier ($O_{id}$), and the value is a pair of recent timestamp ($ts$) and obsolete object counter ($cnt$). When an existing object is deleted or updated, it becomes an obsolete object in the LSM tree, so we update $ts$ to the current timestamp and increment $cnt$ for the UM entry of $O_{id}$. On the contrary, if we perform LSM-aware UM cleanings as in Section~\ref{sec:cleaning}, we decrement $cnt$ of $O_{id}$'s UM entry because the cleaning removes the obsolete objects.

\begin{algorithm}[h]
    \SetKwInOut{Input}{input}
    \SetKwInOut{Output}{output}
    \Input{$O_{id}$: Key of an object \newline
            $val$: Timestamp to be compared}
            
    $ts \gets$ UM[$O_{id}$].ts\;    
    \Repeat{ts.compareAndSet(curr, val) is true}{
        $curr \gets$ ts.get()\;
        \If{$curr \geq$ val}{
            return curr\;
        }
    }
    return val\;
 
    \caption{Compare and If Less then Swap}
    \label{alg:umcccils}
\end{algorithm}   

To support concurrency in UM, first, we need to make $ts$ and $cnt$ in the UM entry atomic integers so that any thread 
%in a system 
can read or write the value without conflict. 
In addition, we introduce a new atomic primitive, \textit{Compare and If Less, then Swap} (CILS, for short) to update the timestamp value of an entry. 
% \jianguo{You can say, we propose/introduce a new primitive termed CILS...But you also need to explain why CAS fails.} \jaewoo{edited and explained below} 
The CILS atomic operation is a variation of the Compare and Swap (CAS) operation~\cite{herlihy1991wait}.
%\walid{Add a reference for CAS here.} \jaewoo{Added}
CAS compares the current value of a memory location with a given value, and then modifies it only if the two values are the same. However, CAS fails in our strategy when multiple threads are trying to update the same memory location of $ts$ and the last thread that succeeds in CAS does not hold the highest value of $ts$. If concurrent threads $th_1$ with $ts=4$ and $th_2$ with $ts=5$ try to update the same memory location and $th_1$ succeeds later than $th_2$,  $ts$ ends with 4, which is not the highest $ts$.
%\walid{The above is unclear why CAS does not work.} \jaewoo{Edited to be clear}
The CILS($m$, $v_{old}$, $v_{new}$) operation, as the name implies, compares $v_{old}$ (i.e., the old value) at $m$ (i.e., a memory location) with $v_{new}$ (i.e., the new value), and modifies $m$ only if $v_{old}$ is less than $v_{new}$. 
%\walid{You need to provide a more formal definition of CILS, e.g., CILS(vold,vnew, ...) or something like that and state what it does exactly}. \jaewoo{Edited}
The reason is that UM always needs to hold a value for the most recent object's timestamp that has to be the highest value (most recent timestamp) among the objects with the same key. 

Algorithm~\ref{alg:umcccils} shows how CILS is implemented on UM. In Line 1, a thread gets the atomic integer variable ($ts$) that will be compared to a given $val$. After the thread gets its value ($curr$) atomically (Line 3), it is compared with $val$. If $curr$ is greater than or equal to $val$, that means another thread has already updated the timestamp with the larger value, so we abort the CILS operation (Lines 4-5). If $curr$ is less than $val$, the thread tries to update $curr$ to $val$ (Line 6). If the CAS operation (in Line 6) fails  (i.e., $ts.compareAndSet()$ is false), then the thread repeats the operations until it succeeds. With CILS, we always keep the largest (the most recent) $ts$ for a given object.    

\begin{algorithm}[h]
    \SetKwInOut{Input}{input}
    \SetKwInOut{Output}{output}
    \Input{$O_{id}$: Key of an object \newline
            $ts$: Timestamp of updated object}
            
    $ret \gets$ UM.putIfAbsent($O_{id}$, ($ts$, 1))\;
    \If{ret is not null}{  
        $curr \gets$ UM.CILS($O_{id}$, ts)\;
        % UM.atomicIncCnt($O_{id}$)\;
        $check \gets$ UM.putIfAbsent($O_{id}$, (curr, 1)) \;
        \If{check is not null}{
            UM.atomicIncCnt($O_{id}$) \;
        }
    }
    
    \caption{Update Memo Update}
    \label{alg:umccupdate}
\end{algorithm}  

As discussed in Section~\ref{sec:dataoperations}, delete/update operations in the LSM RUM-tree require the addition or update of the UM entry. Algorithm~\ref{alg:umccupdate} shows how the UM entry is added or updated during the insert/update operation. Right after insertion of an object into the R-tree component, the algorithm is executed. In Line 1, we call putIfAbsent() on UM with the key $O_{id}$ and the value ($ts$, 1). 
This is for initializing the entry atomically if the given $O_{id}$ does not exist in UM where multiple threads execute the putIfAbsent() function at the same time, only one thread will get $null$ as a return, and the other threads will get some mapping value (i.e., ($ts$, $cnt$)) for $O_{id}$.
If the entry exists, $ret$ is not $null$, and we execute the CILS operation as in Algorithm~\ref{alg:umcccils}, then atomically increment $cnt$ of the entry for $O_{id}$ in UM (Lines 4-6).

For the UM cleaning policies discussed in Section~\ref{sec:cleaning}, we decrement $cnt$ of an UM entry and remove it if the value equals 0. During the cleaning operations, $ts$ of an object $O_{id}$ remains the same, but we only need to decrement $cnt$ for the obsolete objects removed from the LSM RUM-tree (Algorithm~\ref{alg:umccclean}).

\begin{algorithm}[h]
    \SetKwInOut{Input}{input}
    \SetKwInOut{Output}{output}
    \Input{$O_{id}$: Key of an object }
    
    $ret \gets$ UM.atomicDecCnt($O_{id}$)\;
    
    \If{ret == 0}{
        UM.remove($O_{id}$, 0)\;
    }
    
    \caption{Update Memo Clean}
    \label{alg:umccclean}
\end{algorithm}  

\subsection{Implementation and Correctness}
% \jianguo{You didn't explain how to actually implement CLIS.} \jaewoo{stated 3 paragraphs before}
Because there is no hardware support (CPU instruction) for the CILS operation, we need to emulate it in high-level languages using a spin-lock approach with CAS as in Algorithm~\ref{alg:umcccils}. 
Next, we need to use a thread-safe UM data structure for concurrency. As in Algorithm~\ref{alg:umcccils}-\ref{alg:umccclean}, UM operations need to be thread-safe as multiple-threads read and/or write simultaneously. \textit{ConcurrentHashMap}~\cite{concurrenthashmap} in JAVA, 
% \jianguo{add the reference}\jaewoo{added} 
is an example for this purpose that supports thread-safe operations by having separate locks for different hash buckets for various HashMap operations, e.g., put(), putIfAbsent(), or remove(). In addition,  Variables $ts$ and $cnt$ need to support atomic operations for incrementing and decrementing $cnt$ and compareAndSet() on $ts$ in the CILS operation. 

% Next, we show that UM operations are thread-safe on concurrent read and/or write.
To ensure that CILS and UM operations are thread-safe, we divide the operations into three cases that UM can handle, and 
%try to 
we prove 
%its 
their 
correctness.

\subsubsection{Concurrent Updates on UM Entries}
Assume that $UM[O_{id}].ts$ (the current timestamp of an object) is \textbf{1} and $UM[O_{id}].cnt$ (the current number of obsolete objects for $O_{id}$) is \textbf{1}. Assume further that two threads, say $th_{1}$ and $th_{2}$, update UM concurrently with timestamp values \textbf{6} and \textbf{7} as in Algorithm~\ref{alg:umccupdate}. After the execution of Line 1, both $th_{1}$ and $th_{2}$ get the same $ret = 1$. 
%After that, 
Then, 
they execute the CILS operations as in Algorithm~\ref{alg:umcccils}. In Line 3, both threads have the same $curr$, which is equal to \textbf{1}. Thus, both the concurrent executions for $th_{1}$ and $th_{2}$ skip Lines~4 and~5 as $val$ is \textbf{6} and \textbf{7} for $th_{1}$ and $th_{2}$, respectively. Then, each thread executes $ts.compareAndSet(curr, val)$ at Line 6. Because $ts$ is an atomic variable and the $compareAndSet()$ operation atomically sets $val$ only if $curr$ equals $val$, either $th_{1}$ or $th_{2}$ will succeed and exit from CILS. i) If $th_{1}$ succeeds, the current $UM[O_{id}].ts$ will be set to \textbf{6} and exit the CILS operation.
At this point, $th_{2}$ will continue the loop again. Because $curr$ is now set to \textbf{6} and $val$ of $th_{2}$ is \textbf{7}, it skips Lines 4-5 and tries to execute $compareAndSet(curr, val)$ again. After $th_{2}$ completes the operation, $UM[O_{id}].ts$ is set to \textbf{7}, which is the desired result of $ts$ (i.e., the most recent timestamp of $O_{id}$). On the contrary, ii) if $th_{2}$ succeeds the $compareAndSet(curr, val)$ first in the initial loop, then $UM[O_{id}].ts$ would be set to \textbf{7}.
%, directly. 
In this case, $th_{1}$ will exit the CILS in Line~5 of the second loop because $UM[O_{id}].ts = 7$ is already larger than $val$ of $th_{1}$, which is \textbf{6}.
After the CILS operation, both threads execute an atomic increment of $cnt$ for $O_{id}$ and the final value will be \textbf{3} as it is incremented twice by $th_{1}$ and $th_{2}$. 
Even if there are $N$ 
%multiple 
interleaving threads  
concurrently accessing 
the same UM entity, regardless of the $ts$ order, each thread increments $cnt$, and then updates $UM[O_{id}].ts$ only when the $ts$ is less then a thread's $ts$. Therefore, this execution is what is expected to happen and would result in correct behavior of the concurrent updates.
%\walid{Can you then argue given the above that this concurrent interleaving in both cases is correct and is what needs to happen?}\jaewoo{edited}

\subsubsection{Concurrent Cleaning of UM Entries}
Assume that $th_{3}$ performs the Vacuum Cleaning on a leaf node of R-tree for $O_{id}$ while $th_{4}$ executes the Clean-Upon-Flush operation, which also 
%includes 
involves
$O_{id}$. The cleaning process reduces obsolete objects from LSM RUM-tree, and does not affect the $ts$ of a UM entry.
Thus, we only need to execute an atomic decrement of $UM[O_{id}].cnt$ for UM updates. 
Either one of the two threads $th_{3}$ or $th_{4}$ will get $ret = 0$. This will be the thread executed last because of atomicDecCnt(). Hence, only this last thread will remove the UM entry for $O_{id}$. There is an edge case where the removal of an UM entry 
%leads 
could lead
to a wrong result. We handle this case in the next section.

\subsubsection{Concurrent Update and Clean Operations over UM Entries}
Assume that there are two copies of the object $O_{id}$ in LSM RUM-tree (one current and one obsolete), and both are placed in the memory component. Suppose that the values of $ts$ are \textbf{9} and 7 for the recent and obsolete object, respectively. In this case, the current UM entry (i.e., $UM[O_{id}]$) is $\langle id$, \textbf{9}, 1$\rangle$. We need to check the correctness of the concurrency when multiple threads are involved in both UM update and cleaning operations (i.e., $th_{5}$ updates the UM entry at $ts=\textbf{10}$ and $th_{6}$ executes Vacuum Cleaning to remove the obsolete object, concurrently). 

\textbf{i)} If $th_{5}$ reaches Line~4 of Algorithm~\ref{alg:umccupdate} earlier than Line~1 of Algorithm~\ref{alg:umccclean} of $th_{6}$, it is straightforward that the UM entry will be set to $\langle id$, \textbf{10}, \textbf{2}$\rangle$ by $th_{5}$ and then is updated to $\langle id$, \textbf{10}, \textbf{1}$\rangle$ by $th_{6}$. 

Even \textbf{ii)} if $th_{5}$ executes Line~6 after $th_{6}$'s Lines 1-2, the UM entry changes to $\langle id$, \textbf{10}, \textbf{0}$\rangle$ by $th_{6}$, and then $\langle id$, \textbf{10}, \textbf{1}$\rangle$ by $th_{5}$ because $th_{6}$ has not yet  removed the UM entry (Line~3). After $th_{5}$ finishes the update, Line~3 of $th_{6}$ will be aborted because the current $O_{id}.cnt$ is not equal to 0, which is what is expected to happen. 
%\walid{Need to say here, "which is correct", or "which is what is expected to happen", etc.} \jaewoo{Edited}

Lastly, \textbf{iii)} if $th_{6}$ completes Lines 1-3 while $th_{5}$ executes Line~3, the UM entry for $O_{id}$ has already been removed by $th_{6}$. Therefore, $th_{5}$ continues to re-set the UM entry at Line~4. If there are multiple updates, only one thread will initialize the UM entry with $curr$ and the others just increment $cnt$ atomically (Lines 5-6). This is the correct and expected behavior that is consistent with \textbf{i)} and \textbf{ii)}.
%\walid{Need to say here that this is the correct and the expected behavior for correct execution}. \jaewoo{Edited}

\section{Performance Study}\label{sec:performance}

We evaluate 
the LSM RUM-tree along with its insert, delete, update, and search performances. All the experiments are conducted on a machine running Mac OS 10.15.5 on Intel Core i7 with 2.3 GHz, 16 GB memory, and 512 GB SSD. We use three real datasets, Gowalla~\cite{cho2011friendship}, BerlinMOD~\cite{duntgen2009berlinmod}, and ChicagoTaxi~\cite{chicagoTaxi19} with millions of points: Gowalla (6.4m), BerlinMOD (56m) and ChicagoTaxi (15m). For the datasets, there are 107k, 2k, and 5.2k unique keys (Object IDs), respectively, along with their locations over time.
    
    We implement the LSM RUM-tree inside AsterixDB~\cite{asterixdb}, and compare the LSM RUM-tree with the existing Eager and Validation strategies already implemented in AsterixDB. For the LSM RUM-tree, we set the budget of the in-memory R-tree to 256 MB and the page size to 2 KB following~\cite{alsubaiee2014storage}. The merge policy is set to the \textit{prefix} policy with Threshold=5. We augment the UM implementation into the already existing LSM R-tree. To study the effect of only UM, we do not use any optimization, e.g., a bloom filter~\cite{bloom1970space} or range filter~\cite{alsubaiee2015lsm} on the LSM R-tree since they are orthogonal to the focus of this paper. 
    % Experiments run in a single-thread environment, but we expect that UM  will perform better in a multi-threaded environment due to UM's simplicity. We plan to investigate this issue further in the future. 
%    \walid{with single-threaded, how do you test for concurrency?}\jaewoo{This should have been removed. We have concurrency support.}
    
    We use the following notation to refer to the various cleaning strategies. \textit{UM} denotes LSM RUM-tree without any cleaning strategy, \textit{UM}+\textit{(cleaning strategies)} denotes LSM RUM-tree with combinations of cleaning strategies, where F, M, B, and V refer to Clean Upon \textbf{F}lush, Clean Upon \textbf{M}erge, \textbf{B}uffered Cleaning, and \textbf{V}acuum Cleaning, respectively. 

    As in Sections~\ref{sec:bufferedCleaning} and~\ref{sec:vacuumCleaning}, we tune the threshold to get the best overall performance. Lowering the threshold cleans the UM entries and the in-memory tree nodes more frequently, but requires extra computation cost. We empirically set the threshold to 4 and 8 for the update counter of Buffered Cleaning and the global counter of 
    Vacuum Cleaning, respectively.

    \subsection{Update Performance}\label{sec:updatePerformance}
\begin{figure*}[t]
    \centering
    \begin{tikzpicture}
    \pgfplotsset{footnotesize,samples=10}
    \begin{groupplot}[group style = {group name=update group, group size = 3 by 1, horizontal sep = 25pt}, width = 6.3cm, height = 4.7cm]
        \nextgroupplot[ title = {},
            ybar stacked,
            ymin=0,
        	bar width=7pt,
            % 	nodes near coords,
            % enlargelimits=0.15,
            ylabel={seconds},
            symbolic x coords={Eager,Validation, UM, UM+F, UM+M, UM+FM, UM+BV, UM+FMBV},
            legend style = { column sep = 10pt, legend columns = -1, legend to name = grouplegend,},
            xtick=data,
            x tick label style={rotate=90,anchor=east},
            cycle list name=colorbrewer-RYB,]
            \node[text width=6em,inner sep=0pt,anchor=north west] at (rel axis cs: 0.55,1) {\subcaption{Gowalla}\label{fig:updateGowalla}};
            \addplot+[ybar] plot coordinates {(Eager,301.333) (Validation,142.348) 
              (UM,105.341) (UM+F,111.384) (UM+M,98.795) (UM+FM,111.079) (UM+BV,98.054) (UM+FMBV,99.367)};   \addlegendentry{Update}
            \addplot+[ybar] plot coordinates {(Eager,105.43) (Validation,95.977) 
              (UM,104.263) (UM+F,4.898) (UM+M,102.637) (UM+FM,4.854) (UM+BV,2.837) (UM+FMBV,2.699)};        \addlegendentry{Flush}
            \addplot+[ybar] plot coordinates {(Eager,0) (Validation,0) 
              (UM,0) (UM+F,0) (UM+M,0) (UM+FM,0) (UM+BV,0) (UM+FMBV,0)};                                    \addlegendentry{Merge}
            \coordinate (bottom) at (rel axis cs:0,1); 
        \nextgroupplot[title = {},
            ybar stacked,
            ymin=0,
        	bar width=7pt,
            % 	nodes near coords,
            % enlargelimits=0.15,
            % ylabel={seconds},
            symbolic x coords={Eager,Validation, UM, UM+F, UM+M, UM+FM, UM+BV, UM+FMBV},
            xtick=data,
            x tick label style={rotate=90,anchor=east},
            cycle list name=colorbrewer-RYB,]
            \node[text width=7em,inner sep=0pt,anchor=north west] at (rel axis cs: 0.45,1) {\subcaption{BerlinMOD}\label{fig:updateBerlin}};
            \addplot+[ybar] plot coordinates {(Eager,1071.733) (Validation,1392.051) 
              (UM,974.922) (UM+F,974.354) (UM+M,977.398) (UM+FM,980.044) (UM+BV,509.365) (UM+FMBV,515.311)};
            \addplot+[ybar] plot coordinates {(Eager,461.519) (Validation,426.909) 
              (UM,351.908) (UM+F,7.503) (UM+M,354.923) (UM+FM,7.494) (UM+BV,0.011) (UM+FMBV,0.014)};
            \addplot+[ybar] plot coordinates {(Eager,38.031) (Validation,216.883) 
              (UM,144.32) (UM+F,0.142) (UM+M,43.949) (UM+FM,0.064) (UM+BV,0) (UM+FMBV,0)};
        \nextgroupplot[title = {},
            ybar stacked,
            ymin=0,
        	bar width=7pt,
            % 	nodes near coords,
            % enlargelimits=0.15,
            % ylabel={seconds},
            symbolic x coords={Eager,Validation, UM, UM+F, UM+M, UM+FM, UM+BV, UM+FMBV},
            xtick=data,
            x tick label style={rotate=90,anchor=east},
            cycle list name=colorbrewer-RYB,] 
            \node[text width=8em,inner sep=0pt,anchor=north west] at (rel axis cs: 0.45,1) {\subcaption{ChicagoTaxi}\label{fig:updateChicago}};
             \addplot+[ybar] plot coordinates {(Eager,3167.475) (Validation,389.601) 
              (UM,328.912) (UM+F,330.284) (UM+M,326.888) (UM+FM,328.829) (UM+BV,327.831) (UM+FMBV,344.836)};
            \addplot+[ybar] plot coordinates {(Eager,90.853) (Validation,32.626) 
              (UM,32.217) (UM+F,2.152) (UM+M,31.916) (UM+FM,2.213) (UM+BV,6.009) (UM+FMBV,0.72)};
            \addplot+[ybar] plot coordinates {(Eager,6.449) (Validation,51.106) 
              (UM,43.233) (UM+F,0.106) (UM+M,13.009) (UM+FM,0.047) (UM+BV,4.368) (UM+FMBV,0.084)};
    \end{groupplot}
    % \node[text width=6cm,align=center,anchor=north] at (update group c1r1.south) {\captionof{subfigure}{Pressure relative (P/P)\label{fig:updateGowalla}}};
    % \node[text width=6cm,align=center,anchor=north] at (update group c2r1.south) {\captionof{subfigure}{Pressure relative (P/P)\label{fig:updateBerlin}}};
    % \node[text width=6cm,align=center,anchor=north] at (update group c3r1.south) {\captionof{subfigure}{Pressure relative (P/P)\label{fig:updateChicago}}};
    \node at ($(update group c2r1) + (0cm,-3.9cm)$) {\ref{grouplegend}}; 
    \end{tikzpicture}
    \vspace*{-0.6em}
    \caption{Comparisons of Update Performance on LSM R-trees}
    \label{fig:updatePerformance}
\end{figure*}
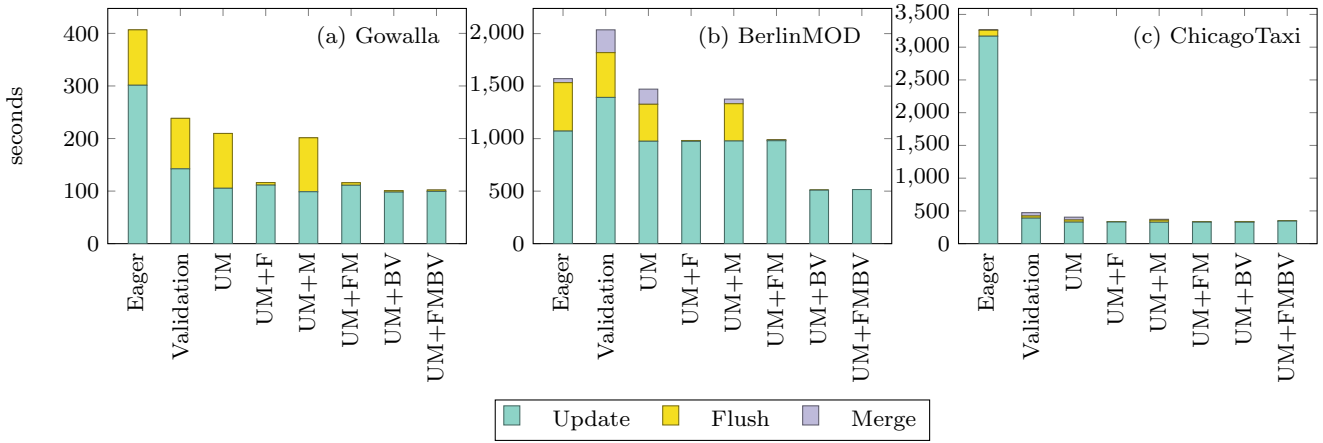        
        To represent various scenarios of update-intensive workloads, first, we sort the data objects by their timestamps, then ingest the data objects into each LSM RUM-tree. The first ingestion of each data object is an \textit{insert}, then is followed by multiple \textit{update} operations. We measure the total execution time to complete the data ingestion of all inserts and updates, excluding setup and data feeding time. When comparing the LSM RUM-tree against the LSM R-Tree, to highlight only the performance of the LSM R-tree, we exclude the point lookup time on primary key in the {\em Eager} strategy of the LSM R-tree, and just feed the old location of an object along with the object delete and update requests.
        
        Figure~\ref{fig:updatePerformance} gives the update performance on three datasets with respect to update procedures, the time to process a flush operation, and the time to process a merge operation by comparing the LSM RUM-tree mechanisms with the {\em Eager} and the {\em Validation} strategies of the LSM R-tree.  As in Section~\ref{sec:cleaning}, the LSM RUM-tree cleaning strategies clean obsolete objects in both the memory and disk layers as well as the UM entries. In update-intensive workloads, cleaning obsolete objects improves the update performance because it avoids unnecessary node splits in the R-tree and also minimizes the flush/merge operations. Table~\ref{table:flushMerge} gives the total number of flush and merge operations for each dataset's updates. From the experiments, observe that the LSM RUM-tree with the various cleaning strategies achieves 3x to 9.6x speedups over the {\em Eager} strategy and 1.4x to 4x better than the {\em Validation}
        strategy depending on the different datasets.
% \vspace*{-0.2em}        
        
        \begin{table}[h]
            \centering
            \begin{adjustbox}{max width=\linewidth}
            \begin{tabular}{ | c | c | c | c | c | c | c |}
            % \begin{tabular}{\linewidth}{| p |*{6}{p{1.5cm}|}}
            \hline
                % strategy & 1 & 2 & 3 \\
                \multirow{2}{*}{strategy} & \multicolumn{2}{c|}{Gowalla} & \multicolumn{2}{c|}{BerlinMOD} & \multicolumn{2}{c|}{ChicagoTaxi}\\
                \cline{2-7}
                & Flush & Merge & Flush & Merge & Flush & Merge\\
                % & F & M & F & M & F & M\\
                % & \shortstack{Flush\\Count} & \shortstack{Merge\\Count} & \shortstack{Flush\\Count} & \shortstack{Merge\\Count} & \shortstack{Flush\\Count} & \shortstack{Merge\\Count}\\
            \hline
                Eager       & 3 & 0 & 28 & 6 & 14 & 3\\ 
                Validation  & 3 & 0 & 30 & 6 & 10 & 2\\ 
                UM          & 3 & 0 & 30 & 6 & 10 & 2\\ 
                UM+F        & 3 & 0 & 30 & 6 & 10 & 2\\ 
                UM+M        & 3 & 0 & 30 & 7 & 10 & 2\\ 
                UM+FM       & 3 & 0 & 30 & 7 & 10 & 2\\ 
                UM+BV       & 0 & 0 & 0  & 0 & 6 & 1\\ 
                UM+FMBV     & 0 & 0 & 0  & 0 & 6 & 1\\ 
            \hline
            \end{tabular}
            \end{adjustbox}
            \vspace*{1em}
            \caption{The number of Flush and Merge operations for each dataset}
            % \vspace*{-1em}
            \label{table:flushMerge}
        \end{table}
        
        In particular, notice the expected cleaning effect of the Clean Upon Flush (i.e., \textit{UM+F}) among all datasets. As it discards obsolete objects on a flush operation, the in-memory R-tree gets much smaller than the original 
        uncleaned 
        one. Thus, the performance of \textit{UM+F} in Figure~\ref{fig:updatePerformance} and Table~\ref{table:flushMerge} reduces the processing time for the flush operation by up to 60x while performing the same number of flush operations. In the update-intensive workloads scenario, the in-memory R-tree contains many obsolete objects. By filtering the obsolete objects in the flushing procedures, the size of flushed component is shrunk and the update operation cost is reduced dramatically. 
        
        Figure~\ref{fig:updateBerlin} illustrates that Clean Upon Merge (i.e., UM+M) significantly reduces the time for the merge operations by a factor of 3 when compared to UM. Notice how efficiently the \textit{UM+M} cleans the obsolete objects in the disk layer. While \textit{UM} merges all objects in the disk layer without cleaning, the \textit{UM+M} discards the obsolete objects using UM  so that the size of merged R-tree is %being 
        much smaller (i.e., resulting in fewer disk I/Os), specifically in update-intensive workloads environment. By reducing the time for the merge operation, %we get 
        UM+M improves the overall performance of the update operations. 
        
        Also, observe the effect of the Buffered and Vacuum Cleaning strategies (i.e., \textit{UM+BV}) 
        as in Figure~\ref{fig:updateBerlin}, where the time for update processing is reduced compared to the other cleaning strategies. While the Gowalla dataset consists of check-in locations of users and the ChicagoTaxi dataset collects pickup locations of taxi trips,  and the BerlinMOD dataset tracks moving objects for each vehicle. As objects in BerlinMOD gradually move in the space covered by the R-tree, the majority of the old and new location
        updates fall in the same R-tree nodes. As \textit{UM+BV} executes an in-memory R-tree cleaning, an unnecessary node split by obsolete objects is avoided. Therefore, the update performance is improved by a factor of 2 in the BerlinMOD dataset. In addition, the in-memory cleaning strategies prevent a flush operation as well. Table~\ref{table:flushMerge} lists the number of flush and merge operations for each dataset. Notice that \textit{UM+BV} reduces the number of flush/merge operations specifically on the moving objects (i.e., BerlinMOD). As in Sections~\ref{sec:bufferedCleaning} and~\ref{sec:vacuumCleaning}, a flush operation is delayed as the cleaning strategies clean the nodes of the in-memory R-tree.
        
        Table~\ref{table:maxsizeUM} lists the maximum size of UM during ingestion. Clearly, the size of UM without any cleaning strategy is bounded to the number of updated objects in the LSM RUM-tree. Observe that the Buffered and Vacuum Cleaning strategies reduce the size of UM up to 92\%
        %\walid{"dramatically" is quite subjective. Be objective and state a number, a quantity, or a percentage, and do not say "dramatically", just say by how much the size gets reduced, etc.}\jaewoo{Edited}
        on the Gowalla and BerlinMOD datasets. Because the ChicagoTaxi is a collection of passengers' pickup points (i.e., random), the effect on the UM size from the Buffered and Vacuum Cleaning strategies is diminished. 

        \begin{table}[h]
            \centering
            \begin{tabular}{ | c | c | c | c |}
            \hline
                strategy & Gowalla & BerlinMOD & ChicagoTaxi \\
            \hline
                UM & 99185 & 2000 & 5221 \\ 
                UM+F & 74334 & 2000 & 5138 \\ 
                UM+M & 99185 & 2000 & 4892 \\ 
                UM+FM & 74334 & 2000 & 4790 \\ 
                UM+BV & 7728 & 156 & 5188 \\ 
                UM+FMBV & 7728 & 156 & 4967 \\ 
            \hline
            \end{tabular}
            \vspace*{1em}
            \caption{Maximum size of UM during update operations}
            \vspace{-1em}
            \label{table:maxsizeUM}
        \end{table}

    \subsection{Search Performance}~\label{sec:queryPerformance}
\begin{figure*}[t]
    \centering
    \begin{tikzpicture}
    \pgfplotsset{footnotesize,samples=10}
    \begin{groupplot}[group style = {group name=query group, group size = 3 by 1, horizontal sep = 25pt}, width = 6.3cm, height = 6cm]
        \nextgroupplot[ title = {},
            ybar=.5pt,
            ymode=log,
            log origin=infty,
        	bar width=1.5pt,
            legend style = { column sep = 6pt, legend columns = -1, legend to name = grouplegend,},
            ylabel={milliseconds},
            xlabel={selectivity},
            x label style={at={(axis description cs:0.5,-0.2)},anchor=north},
            symbolic x coords={.01\%,.07\%, .41\%, 1.56\%, 4.67\%, 11.76\%},
            xtick=data,
            x tick label style={rotate=90,anchor=east},
            cycle list name=colorbrewer-RYB,]
            \node[text width=6em,inner sep=0pt,anchor=north west] at (rel axis cs: 0.01,1) {\subcaption{Gowalla}\label{fig:queryGowalla}};
            % \node [text width=5em,anchor=south west] at (rel axis cs: 0,0)
            %     {\subcaption{Gowalla \label{fig:queryGowalla}}};%<- changed
            % \node[subcaption={Eggs with label fig:a},draw = red,at={(rel axis cs: 0.5, 0.5)}];
            \addplot coordinates {(.01\%,29) (.07\%,88) (.41\%,206) (1.56\%,326) (4.67\%,403) (11.76\%,501)}; \addlegendentry{Eager}
            \addplot coordinates {(.01\%,113) (.07\%,312) (.41\%,705) (1.56\%,1083) (4.67\%,1360) (11.76\%,1735)}; \addlegendentry{Validation}
            \addplot coordinates {(.01\%,31) (.07\%,87) (.41\%,203) (1.56\%,318) (4.67\%,403) (11.76\%,505)}; \addlegendentry{UM}
            \addplot coordinates {(.01\%,6) (.07\%,17) (.41\%,40) (1.56\%,61) (4.67\%,74) (11.76\%,92)}; \addlegendentry{UM+F}
            \addplot coordinates {(.01\%,31) (.07\%,87) (.41\%,203) (1.56\%,315) (4.67\%,404) (11.76\%,505)}; \addlegendentry{UM+M}
            \addplot coordinates {(.01\%,6) (.07\%,18) (.41\%,38) (1.56\%,59) (4.67\%,75) (11.76\%,96)}; \addlegendentry{UM+FM}
            \addplot coordinates {(.01\%,2.1) (.07\%,2.6) (.41\%,5.5) (1.56\%,8.5) (4.67\%,9.9) (11.76\%,12.0)}; \addlegendentry{UM+BV}
            \addplot coordinates {(.01\%,2.3) (.07\%,2.5) (.41\%,6.2) (1.56\%,8.8) (4.67\%,10.1) (11.76\%,12.7)}; \addlegendentry{UM+FMBV}
            \coordinate (bottom) at (rel axis cs:0,1); 
        \nextgroupplot[title = {},
            ybar=.5pt,
            ymode=log,
            log origin=infty,
        	bar width=1.5pt,
        	xlabel={selectivity},
        	x label style={at={(axis description cs:0.5,-0.2)},anchor=north},
        	symbolic x coords={.01\%,.07\%, .41\%, 1.56\%, 4.67\%, 11.76\%},
            xtick=data,
            x tick label style={rotate=90,anchor=east},
            cycle list name=colorbrewer-RYB,]
            \node[text width=6em,inner sep=0pt,anchor=north west] at (rel axis cs: 0.05,1) {\subcaption{BerlinMOD}\label{fig:queryBerlin}};
            \addplot coordinates {(.01\%,3) (.07\%,17) (.41\%,72) (1.56\%,228) (4.67\%,547) (11.76\%,1190)};
            \addplot coordinates {(.01\%,76) (.07\%,235) (.41\%,784) (1.56\%,2303) (4.67\%,5655) (11.76\%,13265)};
            \addplot coordinates {(.01\%,60) (.07\%,179) (.41\%,590) (1.56\%,1612) (4.67\%,3500) (11.76\%,7617)};
            \addplot coordinates {(.01\%,1.1) (.07\%,1.6) (.41\%,4) (1.56\%,6.1) (4.67\%,14.8) (11.76\%,31.2)};
            \addplot coordinates {(.01\%,4.2) (.07\%,3.8) (.41\%,6.4) (1.56\%,19.5) (4.67\%,49.1) (11.76\%,110)};
            \addplot coordinates {(.01\%,3.0) (.07\%,3.1) (.41\%,4.7) (1.56\%,5.8) (4.67\%,12.3) (11.76\%,32.6)};
            \addplot coordinates {(.01\%,0.3) (.07\%,1.3) (.41\%,1.3) (1.56\%,1.4) (4.67\%,1.6) (11.76\%,1.6)};
            \addplot coordinates {(.01\%,0.4) (.07\%,1.4) (.41\%,1.4) (1.56\%,1.5) (4.67\%,1.6) (11.76\%,1.8)};
        \nextgroupplot[title = {},
            ybar=.5pt,
            ymode=log,
            log origin=infty,
        	bar width=1.5pt,
        	xlabel={selectivity},
        	x label style={at={(axis description cs:0.5,-0.2)},anchor=north},
        	symbolic x coords={.01\%,.07\%, .41\%, 1.56\%, 4.67\%, 11.76\%},
            xtick=data,
            x tick label style={rotate=90,anchor=east},
            cycle list name=colorbrewer-RYB,] 
            \node[text width=7em,inner sep=0pt,anchor=north west] at (rel axis cs: 0.03,1) {\subcaption{ChicagoTaxi}\label{fig:queryChicago}};
            \addplot coordinates {(.01\%,16) (.07\%,18) (.41\%,78) (1.56\%,162) (4.67\%,196) (11.76\%,227)};
            \addplot coordinates {(.01\%,349) (.07\%,414) (.41\%,1638) (1.56\%,3245) (4.67\%,4292) (11.76\%,5169)};
            \addplot coordinates {(.01\%,138) (.07\%,180) (.41\%,738) (1.56\%,1466) (4.67\%,1936) (11.76\%,2544)};
            \addplot coordinates {(.01\%,13) (.07\%,15) (.41\%,66) (1.56\%,131) (4.67\%,156) (11.76\%,177)};
            \addplot coordinates {(.01\%,25) (.07\%,29) (.41\%,119) (1.56\%,251) (4.67\%,299) (11.76\%,334)};
            \addplot coordinates {(.01\%,13) (.07\%,15) (.41\%,62) (1.56\%,127) (4.67\%,155) (11.76\%,172)};
            \addplot coordinates {(.01\%,43) (.07\%,46) (.41\%,169) (1.56\%,325) (4.67\%,372) (11.76\%,406)};
            \addplot coordinates {(.01\%,1.2) (.07\%,1.4) (.41\%,5.5) (1.56\%,10.1) (4.67\%,11.9) (11.76\%,13.1)};            
    \end{groupplot}
    % \node[text width=6cm,align=center,anchor=north] at (query group c1r1.south) {\captionof{subfigure}{Pressure relative (P/P)\label{fig:queryGowalla}}};
    % \node[text width=6cm,align=center,anchor=north] at (query group c2r1.south) {\captionof{subfigure}{Pressure relative (P/P)\label{fig:queryBerlin}}};
    % \node[text width=6cm,align=center,anchor=north] at (query group c3r1.south) {\captionof{subfigure}{Pressure relative (P/P)\label{fig:queryChicago}}};
    \node at ($(query group c2r1) + (0,-4.5cm)$) {\ref{grouplegend}}; 
    \end{tikzpicture}
    \vspace*{-0.6em}
    \caption{Comparisons of query performance on LSM R-trees}
    \label{fig:queryPerformance}
\end{figure*}
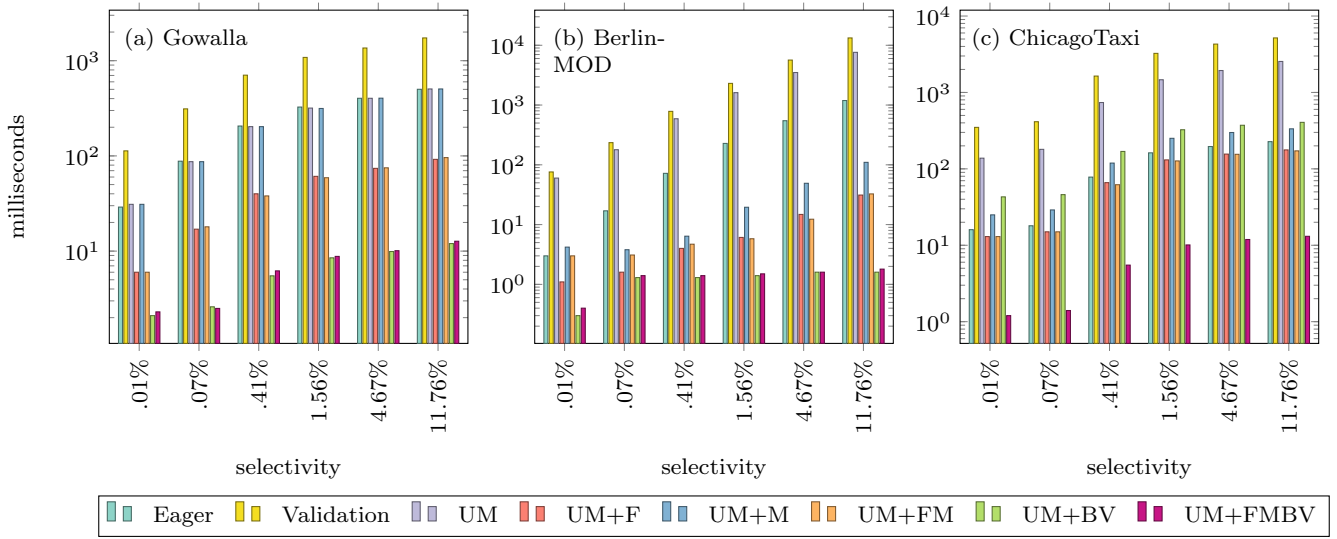

        After data ingestion is completed, we select 100 random query points from each dataset and measure the average time to get the query results. As in~\cite{luo2019efficient}, the {\em Validation} strategy is worse than the {\em Eager} strategy because of the extra validation steps. Figure~\ref{fig:queryPerformance} gives the search (query) performance on each dataset for different query selectivities. Overall, the UM strategy without any cleaning (i.e., \textit{UM}) shows  up to 3x better performance than the {\em Validation} strategy. While  \textit{UM} is comparable to 
        the {\em Eager} strategy on Gowalla (in Figure~\ref{fig:queryGowalla}, it gets worse on the other datasets by one order of magnitude as in Figures~\ref{fig:queryBerlin} and~\ref{fig:queryChicago}). Although \textit{UM} resides in memory, obsolete objects from a query require massive amount of time to validate their states. Therefore, it is essential to have appropriate cleaning strategies as discussed in Section~\ref{sec:cleaning}. Notice that \textit{UM} shows better performance than the {\em Validation} strategy over all datasets because its in-memory structure does not require I/O for validation. 

        The big improvements are due to the cleaning strategies. As discussed in Section~\ref{sec:cleaning}, the cleaning strategies help reduce the size of UM, clean obsolete objects, and shrink the size of the disk component (i.e., the R-trees). First, we focus on the effect of the Clean Upon Flush (i.e., \textit{UM+F}). As  discussed in Section~\ref{sec:cleanUponFlush}, \textit{UM+F} minimizes the number of obsolete objects in the R-tree by filtering them out before the flush operations. Thus, the size of the flushed R-tree into disk is smaller than the original in-memory R-tree. The more the obsolete objects that exist in the in-memory R-tree, the smaller the disk R-tree that gets flushed 
        after being cleaned. 
        As  in Figure~\ref{fig:queryPerformance}, \textit{UM+F} outperforms the {\em Eager} and the {\em Validation} strategies, specifically up to 424x in the BerlinMOD dataset with large query ranges. Because the continuous movements of objects in BerlinMOD makes most of the objects obsolete, the size of the R-trees in disk gets much smaller for \textit{UM+F} and requires fewer disk I/Os for search operations. 
        
        Next, we focus on the advantages of the Buffered  and Vacuum Cleaning strategies (i.e., \textit{UM+BV}) as in Figures~\ref{fig:queryGowalla} and~\ref{fig:queryBerlin}. Due to its ability to clean a node of the in-memory R-tree, the size of the R-tree does not increase significantly in update-intensive workloads, and hence results in minimized flush operations. Therefore, there is no flush/merge on \textit{UM+BV} on the Gowalla and BerlinMOD datasets in Table~\ref{table:flushMerge} and a query operation will not incur disk I/Os. 
        
        Because the update-intensive workloads make most disk objects obsolete, it is likely that the Clean Upon Merge (i.e., \textit{UM+M}) cleans most of the obsolete objects in  disk. When combined with the other strategies, the best performance is \textit{UM+FMBV} as in Figure~\ref{fig:queryChicago}, where among all the cleaning strategies, 
        %achieves 
        achieve
        200x speedup over \textit{UM} on larger selectivities.
        Observe that both the in-memory and disk-side cleaning strategies are important for query performance; in-memory cleaning reduces the obsolete objects (i.e., smaller false-positive query results) and disk-side cleaning leads to fewer I/Os.

   % \subsection{Update/Query Mixed Performance}
   \subsection{Performance of Mixed Update and Query Workloads}
        In a real application, both update and query performance are important to provide stable services. For example, a shared vehicle service would \textit{update} the current location of vehicles and \textit{query} certain ranges to 
        %get a candidate of vehicles 
        identify candidate vehicles
        for a passenger. Another example is a weather emergency application
        that
        would \textit{update} users' locations and \textit{query} 
        %on 
        the emergency area to send an emergency notification to the affected users. For the update/query mixed operations on LSM R-tree, query performance is degraded as the sizes of the in-memory R-tree grow and 
        %the R-tree(s) 
        are flushed into disk.

\begin{figure*}[t]
    \centering
    \begin{tikzpicture}
    \pgfplotsset{footnotesize,samples=10}
    \begin{groupplot}[group style = {group name=uq group, group size = 3 by 1, horizontal sep = 25pt}, width = 6.3cm, height = 5.5cm]
        \nextgroupplot[ title = {},
            axis lines=middle,
            ymin=0,
            xlabel={data processed (\%)},
            x label style={at={(axis description cs:0.5,-0.2)},anchor=north},
            ylabel={seconds},
            y label style={at={(axis description cs:-0.24,.5)},rotate=90,anchor=south},
            enlargelimits = false,
            xticklabels from table={gowallaUpdateQuery.dat}{X},xtick=data,
            legend style = { column sep = 10pt, legend columns = -1, legend to name = grouplegend,},
            ]
            \node[text width=6em,inner sep=0pt,anchor=north west] at (rel axis cs: 0.01,1) {\subcaption{Gowalla}\label{fig:UQGowalla_Update}};
            \addplot[RYB1,thick,mark=*] table [y=eager,x=X]{gowallaUpdateQuery.dat};
            \addlegendentry{Eager}
            \addplot[RYB2,thick,mark=square*] table [y= validation,x=X]{gowallaUpdateQuery.dat};
            \addlegendentry{Validation}]
            \addplot[RYB8,thick,mark=triangle*] table [y= um,x=X]{gowallaUpdateQuery.dat};
            \addlegendentry{UM+FMBV}]
        \nextgroupplot[title = {},
            axis lines=middle,
            ymin=0,
            xlabel={data processed (\%)},
            x label style={at={(axis description cs:0.5,-0.2)},anchor=north},
            y label style={at={(axis description cs:-0.1,.5)},rotate=90,anchor=south},
            enlargelimits = false,
            xticklabels from table={berlinUpdateQuery.dat}{X},xtick=data,
            y tick label style={scaled ticks=base 10:-3},
            legend style={at={(0.02,0.98)},anchor=north west}]
            \node[text width=7em,inner sep=0pt,anchor=north west] at (rel axis cs: 0.01,1) {\subcaption{BerlinMOD}\label{fig:UQBerlinMOD_Update}};
            \addplot[RYB1,thick,mark=*] table [y=eager,x=X]{berlinUpdateQuery.dat};
            % \addlegendentry{Eager}
            \addplot[RYB2,thick,mark=square*] table [y= validation,x=X]{berlinUpdateQuery.dat};
            % \addlegendentry{Validation}]
            \addplot[RYB8,thick,mark=triangle*] table [y= um,x=X]{berlinUpdateQuery.dat};
            % \addlegendentry{UM+FMBV}]
        \nextgroupplot[title = {},
            axis lines=middle,
            ymin=0,
            xlabel={data processed (\%)},
            x label style={at={(axis description cs:0.5,-0.2)},anchor=north},
            y label style={at={(axis description cs:-0.1,.5)},rotate=90,anchor=south},
            enlargelimits = false,
            xticklabels from table={chicagoUpdateQuery.dat}{X},xtick=data,
            y tick label style={scaled ticks=base 10:-3},
            legend style={at={(0.02,0.98)},anchor=north west}]
            \node[text width=7em,inner sep=0pt,anchor=north west] at (rel axis cs: 0.01,1) {\subcaption{ChicagoTaxi}\label{fig:UQChicago_Update}};
            \addplot[RYB1,thick,mark=*] table [y=eager,x=X]{chicagoUpdateQuery.dat};
            % \addlegendentry{Eager}
            \addplot[RYB2,thick,mark=square*] table [y= validation,x=X]{chicagoUpdateQuery.dat};
            % \addlegendentry{Validation}]
            \addplot[RYB8,thick,mark=triangle*] table [y= um,x=X]{chicagoUpdateQuery.dat};
            % \addlegendentry{UM+FMBV}]     
    \end{groupplot}
    \node at ($(uq group c2r1) + (0,-4.0cm)$) {\ref{grouplegend}}; 
    \end{tikzpicture}
    \vspace*{-0.5em}
    \caption{Cumulative update processing time} %\jianguo{What does each line mean? Put the algorithm names.} \jaewoo{added}}
    \label{fig:mixUpdate}
\end{figure*}
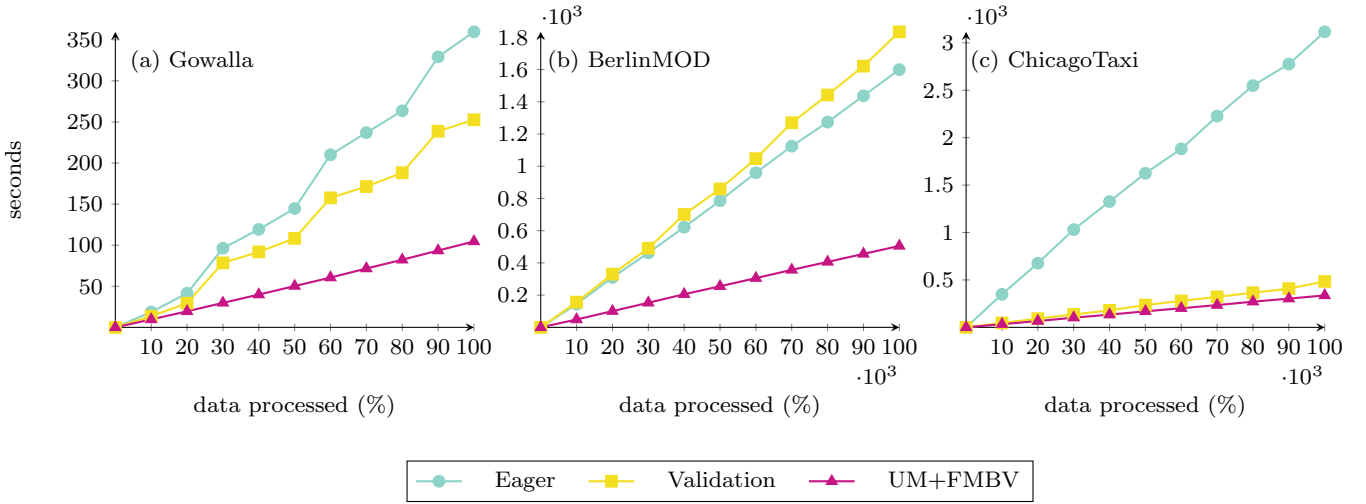    
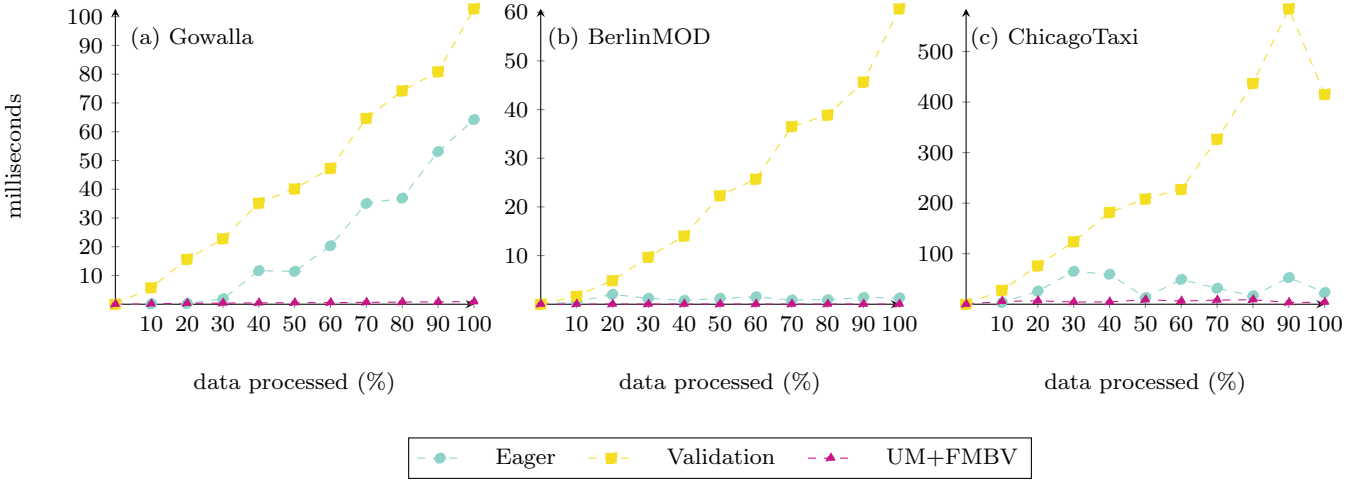
\begin{figure*}[t]
    \centering
    \begin{tikzpicture}
    \pgfplotsset{footnotesize,samples=10}
    \begin{groupplot}[group style = {group name=uq group, group size = 3 by 1, horizontal sep = 25pt}, width = 6.3cm, height = 5.5cm]
        \nextgroupplot[ title = {},
            axis lines=middle,
            ymin=0,
            % ymode=log,
            xlabel={data processed (\%)},
            x label style={at={(axis description cs:0.5,-0.2)},anchor=north},
            ylabel={milliseconds},
            y label style={at={(axis description cs:-0.24,.5)},rotate=90,anchor=south},
            enlargelimits = false,
            xticklabels from table={gowallaAverageQuery.dat}{X},xtick=data,
            legend style = { column sep = 10pt, legend columns = -1, legend to name = grouplegend,},
            ]
            \node[text width=6em,inner sep=0pt,anchor=north west] at (rel axis cs: 0.01,1) {\subcaption{Gowalla}\label{fig:UQGowalla_Query}};
            \addplot[RYB1,dashed,mark=*] table [y=eager,x=X]{gowallaAverageQuery.dat};
            \addlegendentry{Eager}
            \addplot[RYB2,dashed,mark=square*] table [y= validation,x=X]{gowallaAverageQuery.dat};
            \addlegendentry{Validation}]
            \addplot[RYB8,dashed,mark=triangle*] table [y= um,x=X]{gowallaAverageQuery.dat};
            \addlegendentry{UM+FMBV}]
        \nextgroupplot[title = {},
            axis lines=middle,
            ymin=0,
            xlabel={data processed (\%)},
            x label style={at={(axis description cs:0.5,-0.2)},anchor=north},
            y label style={at={(axis description cs:-0.1,.5)},rotate=90,anchor=south},
            enlargelimits = false,
            xticklabels from table={berlinAverageQuery.dat}{X},xtick=data,
            legend style={at={(0.02,0.98)},anchor=north west}]
            \node[text width=7em,inner sep=0pt,anchor=north west] at (rel axis cs: 0.01,1) {\subcaption{BerlinMOD}\label{fig:UQBerlinMOD_Query}};
            \addplot[RYB1,dashed,mark=*] table [y=eager,x=X]{berlinAverageQuery.dat};
            \addplot[RYB2,dashed,mark=square*] table [y= validation,x=X]{berlinAverageQuery.dat};
            \addplot[RYB8,dashed,mark=triangle*] table [y= um,x=X]{berlinAverageQuery.dat};
        \nextgroupplot[title = {},
            axis lines=middle,
            ymin=0,
            xlabel={data processed (\%)},
            x label style={at={(axis description cs:0.5,-0.2)},anchor=north},
            y label style={at={(axis description cs:-0.1,.5)},rotate=90,anchor=south},
            enlargelimits = false,
            xticklabels from table={chicagoAverageQuery.dat}{X},xtick=data,
            legend style={at={(0.02,0.98)},anchor=north west}]
            \node[text width=7em,inner sep=0pt,anchor=north west] at (rel axis cs: 0.01,1) {\subcaption{ChicagoTaxi}\label{fig:UQChicago_Query}};
            \addplot[RYB1,dashed,mark=*] table [y=eager,x=X]{chicagoAverageQuery.dat};
            \addplot[RYB2,dashed,mark=square*] table [y= validation,x=X]{chicagoAverageQuery.dat};
            \addplot[RYB8,dashed,mark=triangle*] table [y= um,x=X]{chicagoAverageQuery.dat};
    \end{groupplot}
    \node at ($(uq group c2r1) + (0,-4.0cm)$) {\ref{grouplegend}}; 
    \end{tikzpicture}
    \vspace*{-0.5em}
    \caption{Averaged query processing time during data ingestions}% \jianguo{What does each line mean? Put the algorithm names.}\jaewoo{added}}
    \label{fig:mixQuery}
\end{figure*}    
\begin{figure*}[t]
    \centering
    \begin{tikzpicture}
    \pgfplotsset{footnotesize,samples=10}
    \begin{groupplot}[group style = {group name=uq group, group size = 3 by 1, horizontal sep = 25pt}, width = 6.3cm, height = 5.5cm]
        \nextgroupplot[ title = {},
            axis lines=middle,
            ymin=0,
            xlabel={data processed (\%)},
            x label style={at={(axis description cs:0.5,-0.2)},anchor=north},
            ylabel={seconds},
            y label style={at={(axis description cs:-0.24,.5)},rotate=90,anchor=south},
            enlargelimits = false,
            xticklabels from table={gowallaCum.dat}{X},xtick=data,
            legend style = { column sep = 10pt, legend columns = -1, legend to name = grouplegend,},
            ]
            \node[text width=6em,inner sep=0pt,anchor=north west] at (rel axis cs: 0.01,1) {\subcaption{Gowalla}\label{fig:UQGowalla_Cum}};
            \addplot[RYB1,thick,mark=*] table [y=eager,x=X]{gowallaCum.dat};
            \addlegendentry{Eager}
            \addplot[RYB2,thick,mark=square*] table [y= validation,x=X]{gowallaCum.dat};
            \addlegendentry{Validation}]
            \addplot[RYB8,thick,mark=triangle*] table [y= um,x=X]{gowallaCum.dat};
            \addlegendentry{UM+FMBV}]
        \nextgroupplot[title = {},
            axis lines=middle,
            ymin=0,
            xlabel={data processed (\%)},
            x label style={at={(axis description cs:0.5,-0.2)},anchor=north},
            y label style={at={(axis description cs:-0.1,.5)},rotate=90,anchor=south},
            enlargelimits = false,
            xticklabels from table={berlinCum.dat}{X},xtick=data,
            legend style={at={(0.02,0.98)},anchor=north west}]
            \node[text width=7em,inner sep=0pt,anchor=north west] at (rel axis cs: 0.01,1) {\subcaption{BerlinMOD}\label{fig:UQBerlinMOD_Cum}};
            \addplot[RYB1,thick,mark=*] table [y=eager,x=X]{berlinCum.dat};
            % \addlegendentry{Eager}
            \addplot[RYB2,thick,mark=square*] table [y= validation,x=X]{berlinCum.dat};
            % \addlegendentry{Validation}]
            \addplot[RYB8,thick,mark=triangle*] table [y= um,x=X]{berlinCum.dat};
            % \addlegendentry{UM+FMBV}]
        \nextgroupplot[title = {},
            axis lines=middle,
            ymin=0,
            xlabel={data processed (\%)},
            x label style={at={(axis description cs:0.5,-0.2)},anchor=north},
            y label style={at={(axis description cs:-0.1,.5)},rotate=90,anchor=south},
            enlargelimits = false,
            xticklabels from table={chicagoCum.dat}{X},xtick=data,
            legend style={at={(0.02,0.98)},anchor=north west}]
            \node[text width=7em,inner sep=0pt,anchor=north west] at (rel axis cs: 0.01,1) {\subcaption{ChicagoTaxi}\label{fig:UQChicago_Cum}};
            \addplot[RYB1,thick,mark=*] table [y=eager,x=X]{chicagoCum.dat};
            % \addlegendentry{Eager}
            \addplot[RYB2,thick,mark=square*] table [y= validation,x=X]{chicagoCum.dat};
            % \addlegendentry{Validation}]
            \addplot[RYB8,thick,mark=triangle*] table [y= um,x=X]{chicagoCum.dat};
            % \addlegendentry{UM+FMBV}]     
    \end{groupplot}
    \node at ($(uq group c2r1) + (0,-4.0cm)$) {\ref{grouplegend}}; 
    \end{tikzpicture}
    \vspace*{-0.5em}
    \caption{Cumulative processing time, where 99\% updates and 1\% queries executed with each dataset}
    \label{fig:mixCum}
\end{figure*}
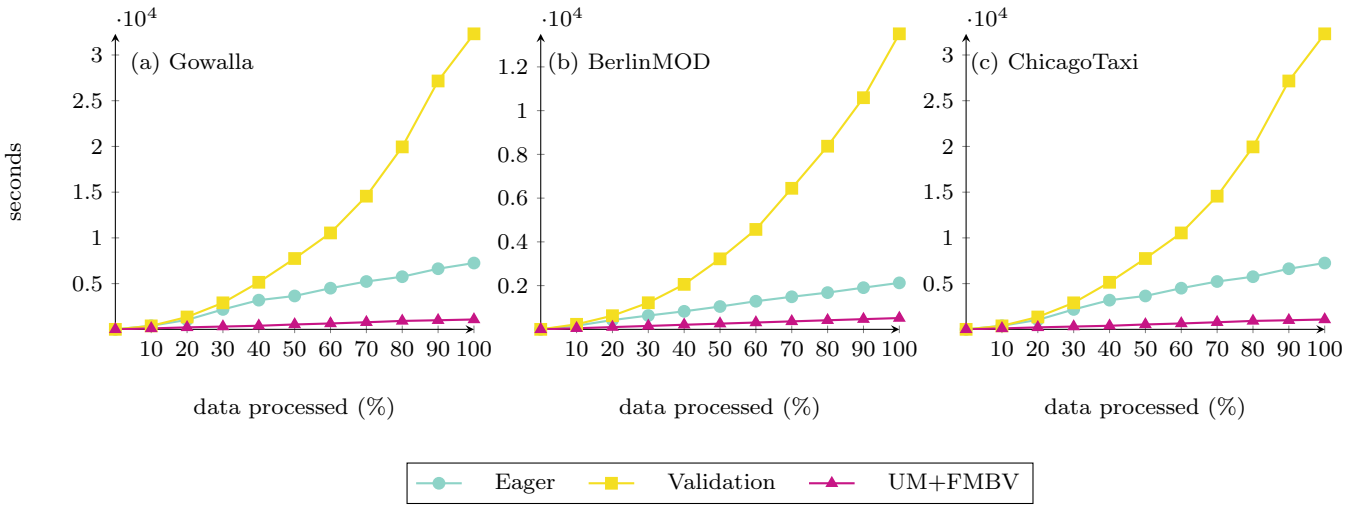

        In the previous experiment in Section~\ref{sec:queryPerformance}, we execute random queries after all of the data ingestion has completed. In this experiment, we use the same three datasets, but 
        %process queries in-between 
        interleave queries with 
        the insert/update operations. To 
        %replicate 
        mimic
        a real update/query scenario using the dataset, we sort the data objects by their timestamps, as explained in Section~\ref{sec:updatePerformance}, then choose 1\% of random objects in each dataset, and use their locations as query points rather than updating the locations of the objects. The range of the query varies uniformly up to 0.01\% of the data space. In other words, Gowalla contains 6.4m objects in total so that there are about 64k (1\%) queries 
        %in-between 
        interleaved with
        6.3m (99\%) updates.  The same applies to the BerlinMOD and ChicagoTaxi datasets. Figures~\ref{fig:mixUpdate},~\ref{fig:mixQuery}, and~\ref{fig:mixCum} give the cumulative update processing time, the average query processing time for each data processing stage, and the cumulative processing time to execute both update and query operations for each dataset. We compare \textit{UM+FMBV} with the {\em Eager} and the {\em Validation} strategies. \textit{UM+FMBV} shows the best performance overall in Sections~\ref{sec:updatePerformance} and~\ref{sec:queryPerformance}. 
%        \walid{is any concurrency being tested in the experiment above? If yes, state so} \jaewoo{We don't have it here, but we have concurrency test in Fig~\ref{fig:multithreadUpdatePerformance}}

        From Figure~\ref{fig:mixCum}, because of the high computational overhead of the validation step, the overall processing time of the {\em Validation} strategy is up to 6.7 times longer
        % much worse
%        \walid{Do not use "much worse" subjective. Say worse by how much} \jaewoo{Edited}
        than \textit{UM+FMBV} that outperforms both the {\em Eager} and the {\em Validation} strategies. The reasons for this difference in performance are as follows. First, the simple update procedures using UM result in improved performance as discussed in Section~\ref{sec:dataoperations}. The update operation in UM is straightforward, and does not incur extra maintenance cost. Second, the flush or merge operation slows down the update processing time. As in Figure~\ref{fig:UQGowalla_Update}, there are three leaps on cumulative update processing time for the {\em Eager} and the {\em Validation} strategy due to the flush operations (i.e., flush executes during 20-30\%, 50-60\%, and 80-90\% of the processed data, respectively). However, \textit{UM+FMBV} does not incur the flush operation with the  Gowalla dataset because of its ability to clean nodes only through the Buffered and Vacuum cleaning strategies.  Also, observe that both cleaning strategies contribute to the improved update performance by making the in-memory R-tree avoid unnecessary node splits from obsolete objects. Even though \textit{UM+FMBV} has flush operations with BerlinMOD (Figure~\ref{fig:UQBerlinMOD_Update}) and ChicagoTaxi (Figure~\ref{fig:UQChicago_Update}), the Clean Upon Flush makes the size of a flushed R-tree very small in update-intensive workloads, and thus the update processing time is stable. Lastly, the search performance under the {\em Eager} and the {\em Validation} strategies is degraded as updates are stacked in the LSM R-tree. The search processing time of the LSM R-tree increases due to the growing sizes of the R-tree(s), disk I/Os, and the calculation overhead of the validation steps.
        
        In Figure~\ref{fig:UQGowalla_Query}, the search performance degrades in all strategies due to updates to the data. Although \textit{UM+FMBV} does not incur disk I/O, the growing size of the R tree in memory increases the search processing time. Also, disk I/Os are critical for search performance as in the Eager and the Validation strategies in Figure~\ref{fig:UQGowalla_Query}. The search processing time increases up to 6x after the flush operations (near the 30\%, 60\%, and 90\% of data processing). Compared to the {\em Eager} and the \textit{UM+FMBV} strategies, the {\em Validation} strategy shows worst search performance during updates, specifically in Figures~\ref{fig:UQBerlinMOD_Query} and~\ref{fig:UQChicago_Query} because of the validation steps using the primary key index. As the BerlinMOD and ChicagoTaxi datasets contain small numbers of distinct objects (2k and 5.2k unique objects, respectively), candidates that result from a search contain many obsolete objects, and  validation using the \textit{primary key index} results in higher computational overheads for the index traversal, and may lead to more disk I/O. In Figure~\ref{fig:UQBerlinMOD_Query}, towards the end of the data processing, \textit{UM+FMBV} (0.025 ms) achieves 52x and 2400x speedup on search processing time over the {\em Eager} (1.3 ms) and the {\em Validation} (60 ms) strategies. With the UM strategy, the memo structure is based on a hash map, and resides only in the memory layer so that the computational overhead of the validation steps is very small. Overall, the update and search performances of the LSM-RUM tree with cleaning strategies significantly outperform the conventional LSM R-tree strategies.

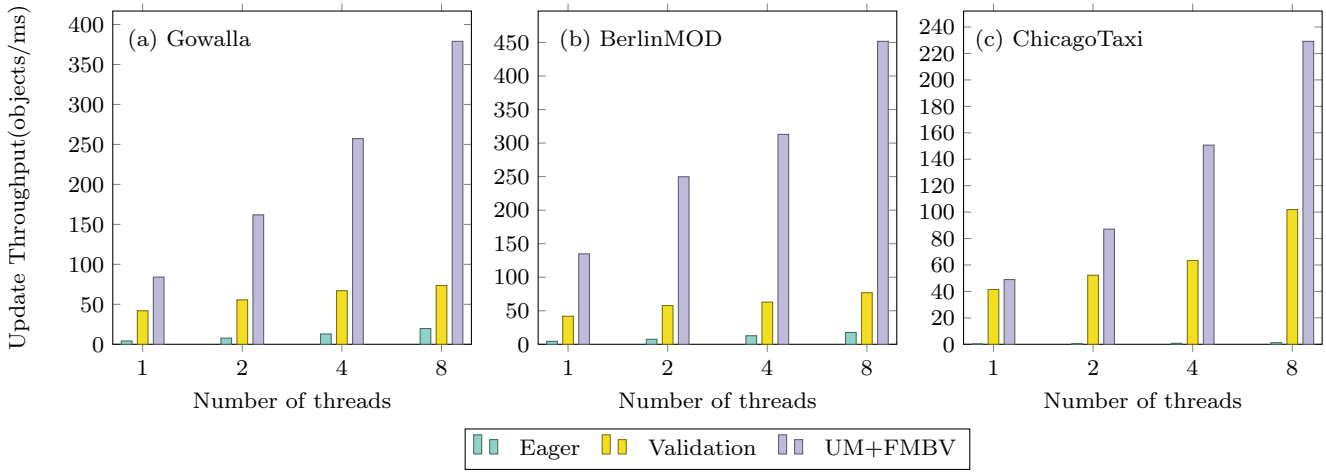
\begin{figure*}[t]
    \centering
    \begin{tikzpicture}
    \pgfplotsset{footnotesize,samples=10}
    \begin{groupplot}[group style = {group name=query group, group size = 3 by 1, horizontal sep = 25pt}, width = 6.3cm, height = 6cm]
        \nextgroupplot[ title = {},
            ybar=2pt,
            % ymode=log,
            % log origin=infty,
            ymin=0,
        	bar width=4pt,
            legend style = { column sep = 6pt, legend columns = -1, legend to name = grouplegend,},
            ylabel={Update Throughput(objects/ms)},
            xlabel={Number of threads},
            x label style={at={(axis description cs:0.5, 0)},anchor=north},
            symbolic x coords={1, 2, 4, 8},
            xtick=data,
            x tick label style={rotate=0,anchor=north},
            cycle list name=colorbrewer-RYB,]
            \node[text width=6em,inner sep=0pt,anchor=north west] at (rel axis cs: 0.01,1) {\subcaption{Gowalla}\label{fig:multithreadUpdateGowalla}};
            % \node [text width=5em,anchor=south west] at (rel axis cs: 0,0)
            %     {\subcaption{Gowalla \label{fig:queryGowalla}}};%<- changed
            % \node[subcaption={Eggs with label fig:a},draw = red,at={(rel axis cs: 0.5, 0.5)}];
            \addplot coordinates {(1,4.141853715) (2,7.733687282) (4,12.91017243) (8,19.74089519)}; \addlegendentry{Eager}
            \addplot coordinates {(1,41.94503255) (2,55.57790085) (4,67.05128685) (8,73.51640289)}; \addlegendentry{Validation}
            \addplot coordinates {(1,84.05862015) (2,161.8905194) (4,257.2780529) (8,378.9411246)}; \addlegendentry{UM+FMBV}
            \coordinate (bottom) at (rel axis cs:0,1); 
        \nextgroupplot[title = {},
            ybar=2pt,
            % ymode=log,
            % log origin=infty,
            ymin=0,
        	bar width=4pt,
            xlabel={Number of threads},
            x label style={at={(axis description cs:0.5, 0)},anchor=north},
        	symbolic x coords={1, 2, 4, 8},
            xtick=data,
            x tick label style={rotate=0,anchor=north},
            cycle list name=colorbrewer-RYB,]
            \node[text width=7em,inner sep=0pt,anchor=north west] at (rel axis cs: 0.05,1) {\subcaption{BerlinMOD}\label{fig:multithreadUpdateBerlin}};
            \addplot coordinates {(1,4.306132912) (2,7.423959296) (4,12.77942772) (8,17.66551897)};
            \addplot coordinates {(1,41.8014196) (2,57.75755078) (4,62.83394828) (8,76.88074911)};
            \addplot coordinates {(1,134.8114079) (2,249.8072615) (4,313.0102831) (8,451.724595)};
        \nextgroupplot[title = {},
            ybar=2pt,
            % ymode=log,
            % log origin=infty,
            ymin=0,
        	bar width=4pt,
            xlabel={Number of threads},
            x label style={at={(axis description cs:0.5, 0)},anchor=north},
        	symbolic x coords={1, 2, 4, 8},
            xtick=data,
            x tick label style={rotate=0,anchor=north},
            cycle list name=colorbrewer-RYB,] 
            \node[text width=7em,inner sep=0pt,anchor=north west] at (rel axis cs: 0.03,1) {\subcaption{ChicagoTaxi}\label{fig:multithreadUpdateChicago}};
            \addplot coordinates {(1,0.200584368) (2,0.396935789) (4,0.722877703) (8,1.210034399)};
            \addplot coordinates {(1,41.44503875) (2,52.23508062) (4,63.40285786) (8,101.9035817)};
            \addplot coordinates {(1,48.97217135) (2,87.11668782) (4,150.6708681) (8,229.1769491)};         
    \end{groupplot}
    \node at ($(query group c2r1) + (0,-3.6cm)$) {\ref{grouplegend}}; 
    \end{tikzpicture}
    % \vspace*{-0.6em}
    \caption{Comparisons of update performance on multi-threads environment}
    \label{fig:multithreadUpdatePerformance}
\end{figure*}
    \subsection{Multi-threads Support}
    As discussed in Section~\ref{sec:ccum}, we provide a mechanism to support concurrency control on the UM structure so that it can also run in a multi-threaded environment. To analyze its performance, we vary the number of threads running the Eager and Validation strategies, and compare the update performance with our LSM RUM-tree with all the cleaning strategies. We use the same real datasets, Gowalla, BerlinMOD, and ChicagoTaxi, and feed the data files for each thread running the update operations. 
    
    %As we can see in
    From
    Figure~\ref{fig:multithreadUpdatePerformance}, the LSM RUM-tree shows superior performance 
    %than 
    over
    the previous strategies. For the Gowalla dataset, the 
    LSM RUM-tree processes 378.9 objects/ms with 8 concurrent working threads, which is 19.9 and 5.2 times more throughput than the eager (73.5 objects/ms) and validation (19.7 objects/ms) strategies, respectively. Also, the speedup achieved from increasing the number of threads from a single thread to 8 threads is 4.5x in the LSM RUM-tree, which is much larger than the one in the validation strategy, where the latter achieves only 1.7x enhancement in performance. The results from Datasets BerlinMOD and ChicagoTaxi are similar.
    %to this. 
    The LSM RUM-tree achieves a throughput of 451.7 and 229.2 objects/ms for the two datasets, respectively. In contrast,  the Validation strategy achieves only 73.5 and 19.7 objects/ms and the Eager strategy achieves 101.9 and 1.2 objects/ms.
    The LSM RUM-tree clearly outperforms both the Validation and the Eager strategies using an LSM R-tree.
    % ~\ref{table:flushMerge}

% \vspace{-0.05in}
\section{Conclusions}\label{sec:conclusion}
    In this paper, we introduce a new index structure, termed the LSM RUM-tree, that leverages an in-memory Update Memo structure (UM) with an LSM-based R-tree to enhance the update and search performances of update-intensive spatial data workloads. We illustrate how to utilize UM in the LSM RUM-tree for delete, update, and search operations. The in-memory UM structure provides efficient validation on query processing as well as simplified update operations. Making UM light-weight is important to be held in memory. To achieve this, we provide four UM cleaning strategies: Buffered Cleaning, Vacuum Cleaning, Clean Upon Flush, and Clean Upon Merge. These strategies not only clean UM entries to shrink its size, but also improve search performance as they also help shrink the size of the disk R-trees, and hence reduce the I/O overheads. The performance study demonstrates that the LSM RUM-tree handles update-intensive workloads efficiently, and outperforms the {\em Eager} and {\em Validation} strategies in the LSM R-tree. Also, the search performance is improved.
    We plan to expand this research to investigate how updates memos can enhance the performance of other secondary indexes beyond R-trees (e.g., B$^+$-trees as a secondary index or inverted indexes) for update-intensive workloads. 
    % Another interesting direction is to investigate the UM technique in multi-threaded environments and it is expected to achieve enhanced performance due to its simplicity. 

\begin{acknowledgements}
Walid G. Aref acknowledges the support of the National Science Foundation under Grant Number IIS-1910216.
\end{acknowledgements}

\bibliographystyle{spmpsci}      % mathematics and physical sciences
\def\UrlBreaks{\do\/\do-}
\bibliography{ref} 

% % Non-BibTeX users please use
% \begin{thebibliography}{}
% %
% % and use \bibitem to create references. Consult the Instructions
% % for authors for reference list style.
% % Format for Journal Reference
% Author, Article title, Journal, Volume, page numbers (year)
% Author, Book title, page numbers. Publisher, place (year)
% \end{thebibliography}

\end{document}